\documentclass[preprint,
  onecolumn,
  notitlepage,
  amsmath,amssymb,
  aip,
  apl,
]{revtex4-1}

\usepackage{graphicx}
\usepackage{dcolumn}
\usepackage{bm}
\usepackage{hyperref}
\usepackage{comment}
\usepackage[version=4]{mhchem}
\usepackage{color}

\usepackage{siunitx}
\usepackage[english]{babel}
\usepackage{esdiff}

\begin{document}

\noindent{\sf \LARGE A ballistic graphene superconducting microwave circuit}

\vspace{0.5em}

\noindent Felix E. Schmidt$^{1*}$, Mark D. Jenkins$^{1*}$, Kenji Watanabe$^2$, Takashi Taniguchi$^2$ \& Gary A. Steele$^{1\ddagger}$

\vspace{1em}

{\noindent\em $^1$Kavli Institute of NanoScience, Delft University of
	Technology, PO Box 5046, 2600 GA, Delft, The Netherlands.}
{\\\em $^2$ National Institute for Materials Science, 1-1 Namiki, Tsukuba, 305-0044, Japan.}
{\\\em $^*$ These authors contributed equally to this work.}
{\\\em $^\ddagger$ Email: g.a.steele@tudelft.nl}

\vspace{1em}

{\bfseries

\noindent Josephson junctions (JJ) are a fundamental component of microwave quantum circuits, such as tunable cavities, qubits and parametric amplifiers.
Recently developed encapsulated graphene JJs, with supercurrents extending over micron distance scales, have exciting potential applications as a new building block for quantum circuits. 
Despite this, the microwave performance of this technology has not been explored.
Here, we demonstrate a microwave circuit based on a ballistic graphene JJ embedded in a superconducting cavity.
We directly observe a gate-tunable Josephson inductance through the resonance frequency of the device and, using a detailed RF model, we extract this inductance quantitatively. 
We also observe the microwave losses of the device, and translate this into sub-gap resistances of the junction at $\mu$eV energy scales, not accessible in DC measurements.
The microwave performance we observe here suggests that graphene Josephson junctions are a feasible platform for implementing coherent quantum circuits.
}

\setlength{\parindent}{2em}

\newpage

\section{Introduction}

\noindent The development of ultra-high mobility graphene with induced superconductivity has led to ballistic transport of Cooper pairs over micron scale lengths, supercurrents that persist at large magnetic fields and devices with strongly non-sinusoidal current-phase relations \cite{calado_ballistic_2015a,benshalom_quantum_2015,lee_ultimately_2015,novoselov_roadmap_2012,walsh_graphenebased_2017}
While most measurements of such graphene Josephson junctions (gJJ) have been limited to the DC regime, Josephson junctions in general also play fundamental role in microwave circuits and devices such as qubits or quantum-limited amplifiers \cite{martinis_course_2004,castellanos-beltran_development_2010}.

In these microwave applications, the Josephson junctions used are almost exclusively based on double-angle evaporated aluminum-aluminum oxide tunnel junctions (AlOx)\cite{dolan_offset_1977a}, resulting in amorphous superconductor-insulator-superconductor (SIS) barriers.
Thus far, despite its robust and tunable superconductivity, graphene has not been implemented in this kind of microwave circuitry.
Apart from potentially addressing some of the design and stability issues with AlOx junctions \cite{zeng_direct_2015, zeng_atomic_2016a}, the use of gJJs in such circuits has the additional feature of allowing tunability of the junction properties through an electrostatic gate \cite{calado_ballistic_2015a,benshalom_quantum_2015,lee_ultimately_2015,larsen_semiconductornanowirebased_2015,delange_realization_2015}.
This feature can help address problems like on-chip heating and crosstalk in superconducting circuits where SQUIDs are used as tuning elements.\cite{schreier_suppressing_2008a,sandberg_tuning_2008}.

Here, we present a superconducting microwave circuit based on a ballistic graphene JJ.
The design of our device is such that it also allows DC access to the junction, allowing us to directly compare the DC and RF response of our circuit.
While the gate-tunability enables us to directly tune the resonance frequency of the hybrid gJJ-resonator circuit, we also use the RF response to obtain additional information about the junction typically inaccessible through purely DC characterization.

\section{Results}

\subsection{Circuit description}

\begin{figure}[thb]
	\centering
	\includegraphics[width=\linewidth]{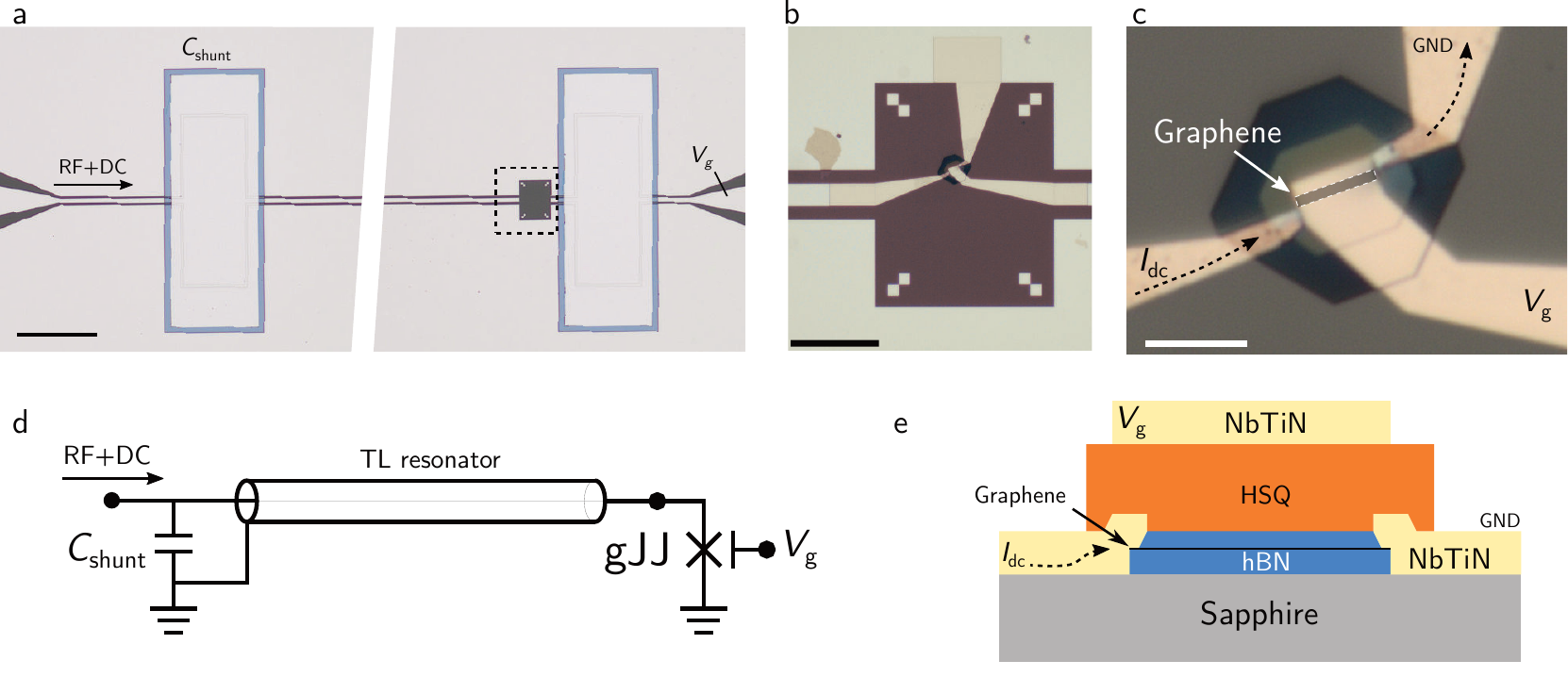}
	\caption[]{\textbf{A gate tunable microwave cavity based on an encapsulated graphene Josephson junction. a,}
		Optical micrograph of the microwave cavity before placing the hBN/G/hBN stack.
		Bright areas are MoRe, dark areas are sapphire substrate.
		Grey area around the parallel plate capacitors is the \ce{Si3N4} shunt dielectric.
		Scale bar \SI{200}{\micro\meter}
		\textbf{b,} Optical micrograph of the gJJ. The cavity center line and the ground plane are connected through the gJJ and NbTiN leads.
		The gate line (right) extends over the entire junction.
		Scale bar \SI{40}{\micro\meter}
		\textbf{c,} Close-up of panel (b) with the graphene channel indicated.
		Dark areas are HSQ for gate insulation.
		Scale bar \SI{5}{\micro\meter}
		\textbf{d,} Sketch of the device circuit.
		The input signals are filtered and merged using a bias tee before being fed on to the feedline (see Methods section and Supplementary Fig. 1).
		\textbf{e,} Schematic cross-section of the gJJ with top-gate, not to scale.}
	\label{fig:figure1}
\end{figure}

The device presented here (Fig.\ref{fig:figure1}) consists of a galvanically accessible graphene Josephson junction embedded in a superconducting coplanar waveguide cavity.
The cavity superconductor is a molybdenum-rhenium (MoRe) alloy sputter-deposited on a sapphire substrate (Fig.\ref{fig:figure1}a).
The coupling to the external feedline is provided by a parallel plate shunt capacitor that acts as semi-transparent microwave mirror \cite{bosman_broadband_2015,singh_molybdenumrhenium_2014}.
In contrast to series capacitors often used as mirrors, the use of shunt capacitors allows us to probe the circuit with steady-state voltages and currents, enabling DC characterization of the gJJ.
A circuit schematic of the device setup is depicted in Fig. \ref{fig:figure1}(d).
The gJJ is made from a graphene and hexagonal boron nitride (BN/G/BN) trilayer stack with self-aligned side contacts \cite{pizzocchero_hot_2016,wang_onedimensional_2013} using a sputtered superconducting niobium titanium nitride (NbTiN) alloy.
The stack is shaped into a junction of length $L=\SI{500}{nm}$ and width $W=\SI{5}{\micro m}$.
Here, $L$ and $W$ denote the distance between the superconducting contacts and lateral extension, respectively.
In order to tune the carrier density of the gJJ, a local DC gate electrode covers the junction and contact area.
Optical micrographs of the device are shown in Figs.\ref{fig:figure1}(b,c) and a schematic cross-section of the gJJ is shown in Fig.\ref{fig:figure1}(e).
Measurements of a similar second device can be found in Supplementary Figs. 10 and 11.

\subsection{DC characterization}

To compare our device with state-of-the-art gJJs, we first perform a purely DC characterization.
We sweep the current-bias ($I_{\rm dc}$) and measure the voltage across the gJJ for different applied gate voltages ($V_\textrm{g}$).
The resulting differential resistance is plotted in Fig. \ref{fig:figure2}(a) and clearly shows a superconducting branch that is tunable through $V_\textrm{g}$.
The junction exhibits $I_\textrm{c}$ in the range of \SI{150}{nA} to \SI{7}{\micro A} for $\lvert V_\mathrm{G} \rvert<\SI{30}{V}$ with significantly lower $I_\textrm{c}$ for $V_\textrm{g}<0$ (p-doped regime) compared to $V_\textrm{g}>0$ (n-doped regime).
Comparing the bulk superconducting gap of our NbTiN leads with the junction Thouless energy, $\Delta/E_\textrm{th}\approx1.52>1$, our device is found to be in the intermediate to long junction regime (see Supplementary Note 7 and Supplementary Figs. 12, 15 and 16).

While in the non-superconducting state (current bias far above the junction critical current $I_\textrm{c}$), the graphene junction shows a narrow peak in its normal resistance associated with low disorder at the charge neutrality point (CNP, at $V_\textrm{g}\approx\SI{-2}{V}$, see Fig.\ref{fig:figure2}(b)), indicating high sample quality.
Some hysteresis in the switching and retrapping currents can also be observed in the measurement (see Supplementary Note 6 for discussion).
We furthermore observe oscillations in both the normal state resistance $R_\textrm{n}$ and the switching and retrapping currents as a function of gate voltage for p-doping of the channel.
We attribute these effects to the presence of PN junctions that form near the graphene-NbTiN contact.
Each of the two NbTiN leads n-dopes the graphene near the respective contact while the main sheet is p-doped by the gate.
The pair of PN junctions produce Fabry-P\'erot interference effects that give rise to the observed oscillations in $I_\textrm{c}$ and $R_\textrm{n}$.
The characteristics of these oscillations indicate that our junction is in the ballistic regime \cite{liang_fabry_2001,miao_phasecoherent_2007,young_quantum_2009,cho_massless_2011,wu_quantum_2012,campos_quantum_2012,rickhaus_ballistic_2013,benshalom_quantum_2015,calado_ballistic_2015a,amet_supercurrent_2016a,borzenets_ballistic_2016a,allen_observation_2017,zhu_supercurrent_2018}.

\subsection{Microwave characterization}

Having established the DC properties of our junction, we turn to the microwave response of the circuit.
Using a vector network analyser, we sweep a microwave tone in the 4 to \SI{8.5}{GHz} range and measure the reflection signal $S_{11}$ of the device for different applied gate voltages $\lvert V_\textrm{g} \rvert \leq \SI{30}{V}$.
The input powers and attenuation used correspond to an estimated intra-cavity photon number of at most 10-20 depending on operating frequency and linewidth.
Further tests were performed at lower powers (down to approximately 0.02 intra-cavity photons) with negligible changes to the cavity line shape and width.
More information on the measurement setup can be found in the Methods section and a detailed sketch in Supplementary Fig. 1.
Figure \ref{fig:figure2}(c) shows the resulting $\lvert S_{11}\rvert$.
A clear resonance dip associated to our device can be tracked as a function of applied gate.
The device exhibits a continuously tunable resonance frequency from \SI{7.1}{GHz} to \SI{8.2}{GHz} with higher frequencies at larger values of $|V_{\rm g}|$.

\subsection{Josephson inductance of the gJJ}

\begin{figure}[thb]
	\centering
	\includegraphics[width=\linewidth]{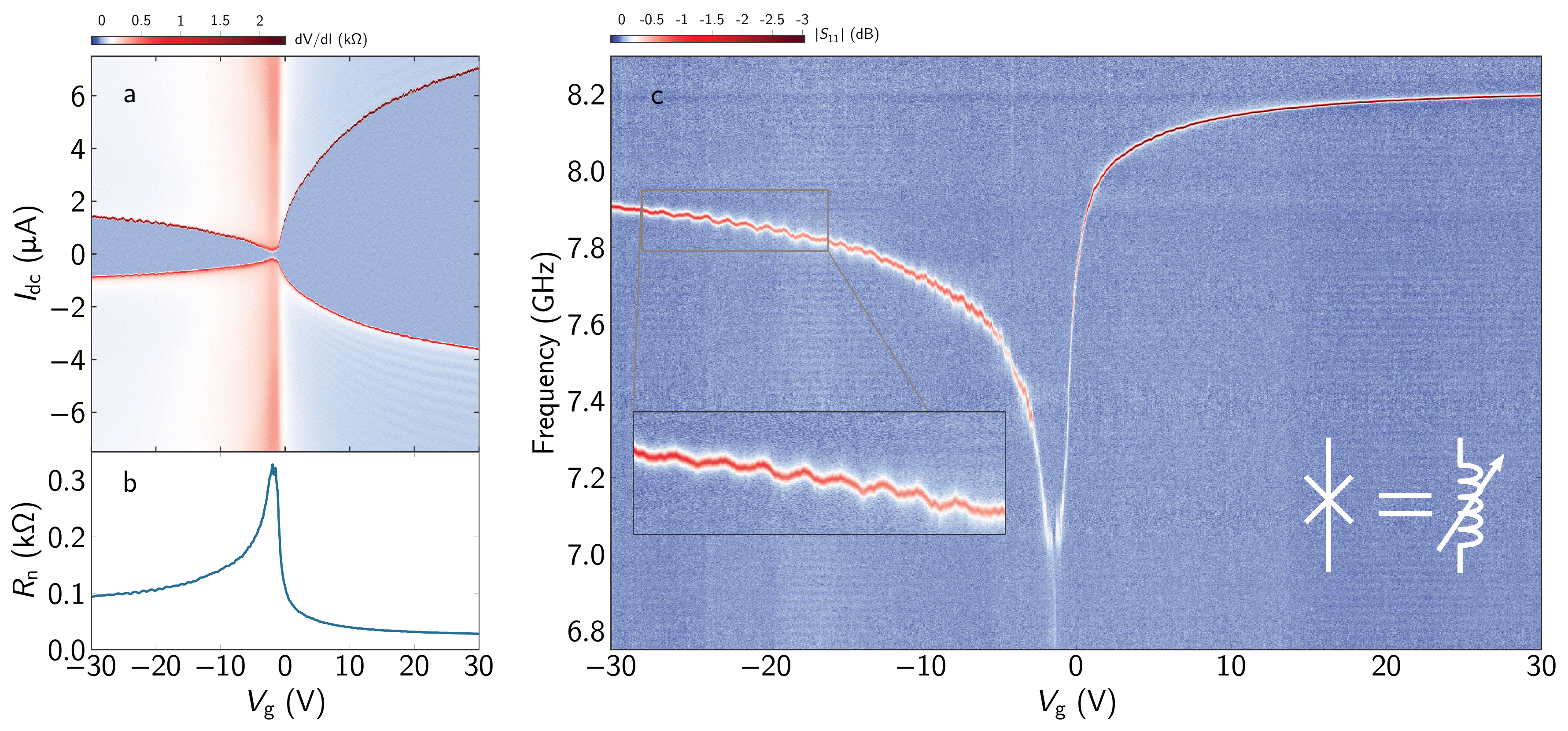}
	\caption[Observation of the Josephson inductance of a ballistic graphene superconducting junction]{\textbf{Observation of the Josephson inductance of a ballistic graphene superconducting junction. a,} Differential resistance across the gJJ for a wide gate voltage range.
		Dark blue denotes area of zero resistance.
		The device shows signatures of FP oscillations on the p-doped side.
		\textbf{b,} Normal state resistance of the gJJ versus gate voltage.
		\textbf{c,} Microwave spectroscopy of the device in the superconducting state versus gate voltage, plotted as the amplitude of the reflection coefficient $\lvert S_{11} \rvert$ after background subtraction.
		The graphene junction acts as a tunable inductor in the microwave circuit, resulting in a cavity frequency that is tuned with gate voltage.
		Inset: The resonance frequency oscillates in phase with the oscillations in \textbf{(a)} and \textbf{(b)}.}
	\label{fig:figure2}
\end{figure}

The origin of the tunable circuit resonance frequency is the variable Josephson inductance of the graphene Josephson junction.
The microwave response of a JJ can be modelled for small currents using an inductor with its Josephson inductance given by:
\begin{equation}
  L_\textrm{j} = \frac{\Phi_0}{2\pi} \left(\diff{I}{\phi}\right)^{-1}, \label{eq:Lj}
\end{equation}
where $\Phi_0$ is the flux quantum.
$L_\textrm{j}$ depends on the superconducting phase difference $\phi$ across the junction and on the derivative of the current-phase relation (CPR).
For small microwave excitations around zero phase ($\phi\simeq 0$) and assuming a sinusoidal CPR, $I=I_\textrm{c}\sin\phi$, this derivative is ${\rm d}I/{\rm d}\phi = I_\textrm{c}$.
This leads to an inductance $L_\textrm{j} = L_{\textrm{J}0} \equiv \frac{\Phi_0}{2\pi I_\textrm{c}}$ which can be tuned by changing the critical current of the junction.
In the device presented here, this junction inductance is connected at the end of the cavity.
When this inductance is tuned, it changes the boundary conditions for the cavity modes and hence tunes the device resonance frequency.
The effect can be illustrated by taking two extreme values of $L_\textrm{j}$ (see Supplementary Fig. 3):
If $L_\textrm{j}\rightarrow 0$ (i.e. $I_\textrm{c}\rightarrow\infty$), the cavity boundary conditions are such that it is a $\lambda/2$ resonator with voltage nodes at both ends.
If, on the other hand $L_\textrm{j}\rightarrow \infty$ ($I_\textrm{c}\rightarrow 0$), the cavity will transition into a $\lambda/4$ resonator with opposite boundary conditions at each end (a voltage node at the shunt capacitor and a current node at the junction end).
This leads to a fundamental mode frequency of about half that of the previous case.
Any intermediate inductance value lies between these two extremes.
Due to the inverse relationship between $I_\textrm{c}$ and $L_\textrm{j}$, the resonance frequency changes very quickly in certain gate voltage regions, having a tuning rate of up to $\mathrm{d}f_0/\mathrm{d}V_\textrm{g}=\SI{1.8}{GHz.V^{-1}}$ at $V_\textrm{g}=\SI{-0.54}{V}$.
This slope could potentially be further increased by increasing the gate lever arm, for example by choosing a thinner gate dielectric.
We again note that the resonance frequency does not saturate within the measured range although the tuning rate at $\vert V_\textrm{g} \rvert=\SI{30}{V}$ is much lower.
Additionally, by comparing Figs. \ref{fig:figure2}(a) and \ref{fig:figure2}(c), we can observe features in the RF measurements that are also present in the DC response.
In particular, the Fabry-P\'erot (FP) oscillations of $I_\textrm{c}$ and $R_\textrm{n}$ seen in the DC measurements result in a modulation of $L_\textrm{j}$, producing corresponding oscillations in the cavity frequency.
By analysing the oscillation period in reciprocal space, we extract a FP cavity length of $L_\textrm{c}\approx \SI{390}{nm}$ (see Supplementary Figs. 13 and 14).
We can thus take $L_\textrm{c}$ as a lower bound for the free momentum scattering and the phase coherence lengths, i.e. $l_{\rm mfp},\xi>L_\textrm{c}$.

\begin{figure}[thb]
	\centering
	\includegraphics[width=\linewidth]{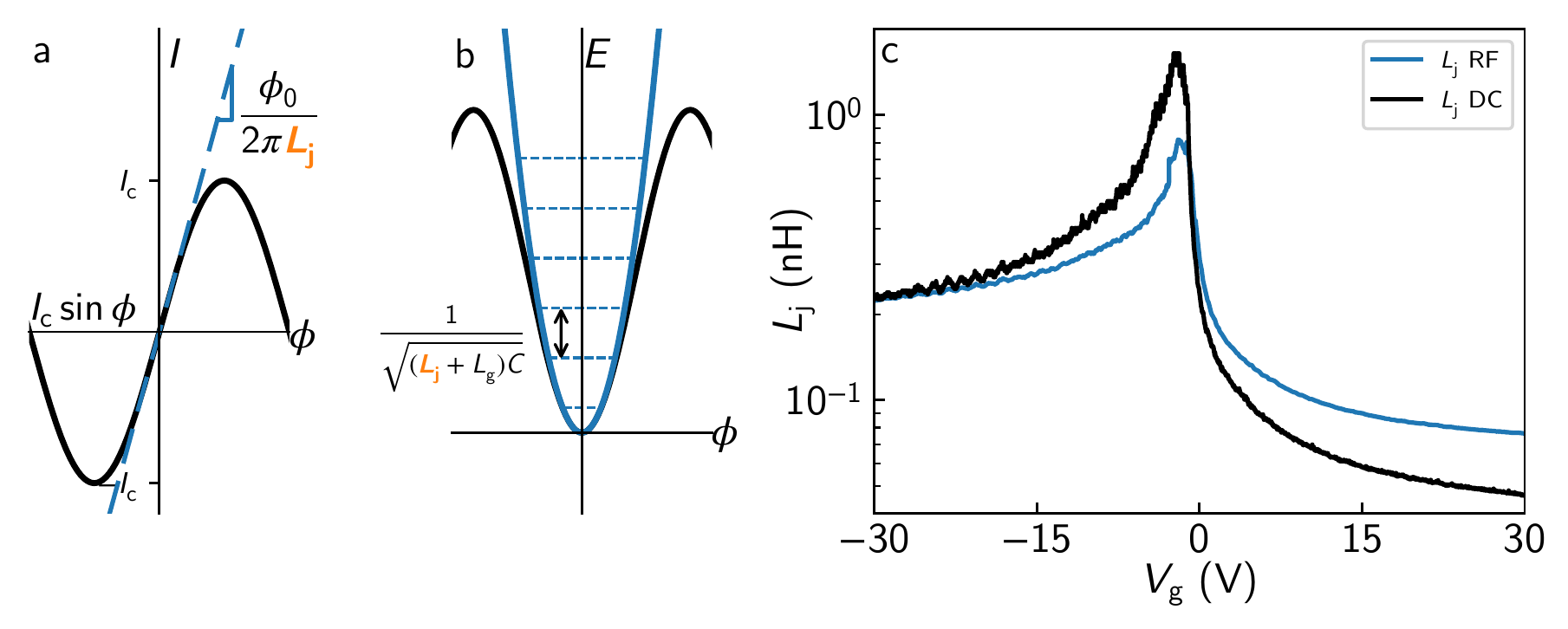}
	\caption[]{\textbf{Josephson inductance extracted from RF and DC measurements. a,}
		Schematic representation of $L_\textrm{j}$ and its relation to the CPR of a Josephson junction.
		$L_\textrm{j}$ can be understood as the slope of the current-phase relation around zero phase bias.
		\textbf{b,} Schematic representation of $L_\textrm{j}$ extraction from the cavity resonance frequency.
		The potential energy near $\phi = 0$ is harmonic, with the fundamental frequency given by the junction inductance $L_\textrm{j}$ and the cavity capacitance $C$ and inductance $L_\textrm{g}$ as $\omega = 1/\sqrt{(L_\textrm{j}+L_\textrm{g})C}$.
		\textbf{c,} Comparison of Josephson inductance $L_\textrm{j}$ extracted from DC measurements (black) and from the microwave measurements (blue).
		We attribute differences to deviations from a sinusoidal current phase relation (see main text for details).
		The error band from our fit of $L_\textrm{j}$ can be found in Supplementary Fig. 4.
	}
	\label{fig:figure3}
\end{figure}

Further analysis of the data presented in Fig.\ref{fig:figure2}c can be used to perform a more quantitative analysis of the Josephson inductance of the gJJ as a function of gate voltage.
As illustrated in Fig. \ref{fig:figure3}(a) and equation (\ref{eq:Lj}), the Josephson inductance $L_\textrm{j}$ is defined according to the slope of the CPR near $\phi=0$ and sets the Josephson energy scale.
For a given assumed CPR, the inductance can be deduced from a DC measurement of the junction $I_\textrm{c}$.
When measuring the RF response of our device, the current in the junction oscillates with a very low amplitude around $\phi=0$.
This directly probes the CPR slope and the Josephson inductance at zero phase bias.
This inductance $L_\textrm{j}$ combined with the cavity inductance $L_\textrm{g}$ and capacitance $C$ determine the resonance frequency (Fig. \ref{fig:figure3}(b)).
An accurate calibration of the cavity parameters then allows us to extract $L_\textrm{j}$ from our measured resonance frequency without assuming any specific CPR.

To accurately obtain $L_\textrm{j}$ from our measurements, we calibrate the parameters of our RF model of the device using simulations and independent measurements, including effects of the kinetic inductance of the superconductor, the capacitance and inductance of the leads connecting the junction to the cavity, and the coupling to the external measurement circuit (see Supplementary Notes 1 and 2, Supplementary Fig. 2 and Supplementary Table 1) leaving only the junction characteristics as the remaining fit parameters.
By fitting the microwave response of the circuit, we obtain the resonance frequency as well as internal and external Q-factors voltage.
Using the model, we then translate this into an extracted inductance $L_\textrm{j}$ of the junction for each gate voltage.

Figure \ref{fig:figure3}c shows the resulting $L_\textrm{j}$ obtained from the dataset in Fig. \ref{fig:figure2}c compared to that obtained by assuming a sinusoidal CPR together with the DC switching currents from Fig. \ref{fig:figure2}a.
At low negative gate voltages we find excellent agreement between the DC and RF models.
As the gate voltage approaches the CNP, we observe clear differences, as the DC value of $L_\textrm{j}$ from a sinusoidal CPR overestimates the inductance obtained from the RF measurements.
For positive gate voltages, on the other hand, the DC value lies well below the one from our microwave measurements.

To understand the implications of these results, we start first with the p-doped regime. 
Since the gJJ is intermediate to long junction regime  and has low contact transparency at high p-doping due to PN junctions at the contacts, it is expected to have a sinusoidal CPR.
In this case, the DC values of $I_\textrm{c}$ should correctly predict Josephson inductance.
The clear agreement between the RF and DC values for $L_\textrm{j}$ in this regime is remarkable, and suggests that we have an accurate RF model of the circuit that can be used to extract direct information about the nature of our junction.
For high n-doping, the DC measurement yields much lower values of $L_\textrm{j}$ than the ones obtained from our RF measurements.
This is in agreement with the fact that high transparency and doping has been observed to produce forward skewing in gJJ CPR \cite{nanda_currentphase_2017} which leads to an underestimation of $L_\textrm{j}$ if a sinusoidal CPR is used in the DC calculation.
On the other hand, the origin of the mismatch for $V_\textrm{g}$ around the CNP is unclear.
Although noise in the bias current can cause DC measurements to overestimate $L_\textrm{j}$, the noise present in our setup cannot account for this deviation.
Alternatively, using the same logic as in the high n-doping case, this deviation could be accounted for with a backward skewed CPR.
However, this is contrary to what has been reported in previous measurements on graphene \cite{english_observation_2016}.

\subsection{Microwave losses in the gJJ}

While tracking the resonance frequency as a function of gate voltage enables us to extract the Josephson inductance, the resonance linewidth provides information about the microwave losses of the gJJ.
The resonance linewidth is also observed to depend on the gate voltage, with minimum values of $\Gamma\sim\SI{2}{MHz}$ at high $|V_{\rm g}|$ and a maximum of \SI{80}{MHz} near the CNP.
We use measurements of an identical circuit without the graphene junction as a benchmark to calibrate the internal and external cavity linewidths.
Using this benchmark together with a model for the junction losses, we find the correct combination of junction parameters that provide the observed frequency and cavity linewidth.
This allows us to quantify the amount of microwave losses attributable to the junction.

We describe the junction using the Resistively Capacitively Shunted Junction (RCSJ) model where the losses are parametrized by a dissipative element $R_\textrm{j}$.
For voltages larger than the superconducting gap $\Delta$ the effective resistance $R_\textrm{j} = R_\textrm{n}$ is that of normal state graphene.
The RF currents applied in our experiment, however, are well below $I_\textrm{c}$, and the associated voltages are also well below the bulk superconducting gap.
In this regime, the correct shunt resistance for the RCSJ model is not the normal state resistance $R_\textrm{n}$ but instead given by the zero-bias sub-gap resistance $R_\textrm{j} = R_{\rm sg}$.
This quantity, which ultimately determines the junction performance in microwave circuits, has not been observed before in graphene as it is only accessible through sub-microvolt excitations, which are difficult to achieve in DC measurements.

\begin{figure}[thb]
	\centering
	\includegraphics[width=\linewidth]{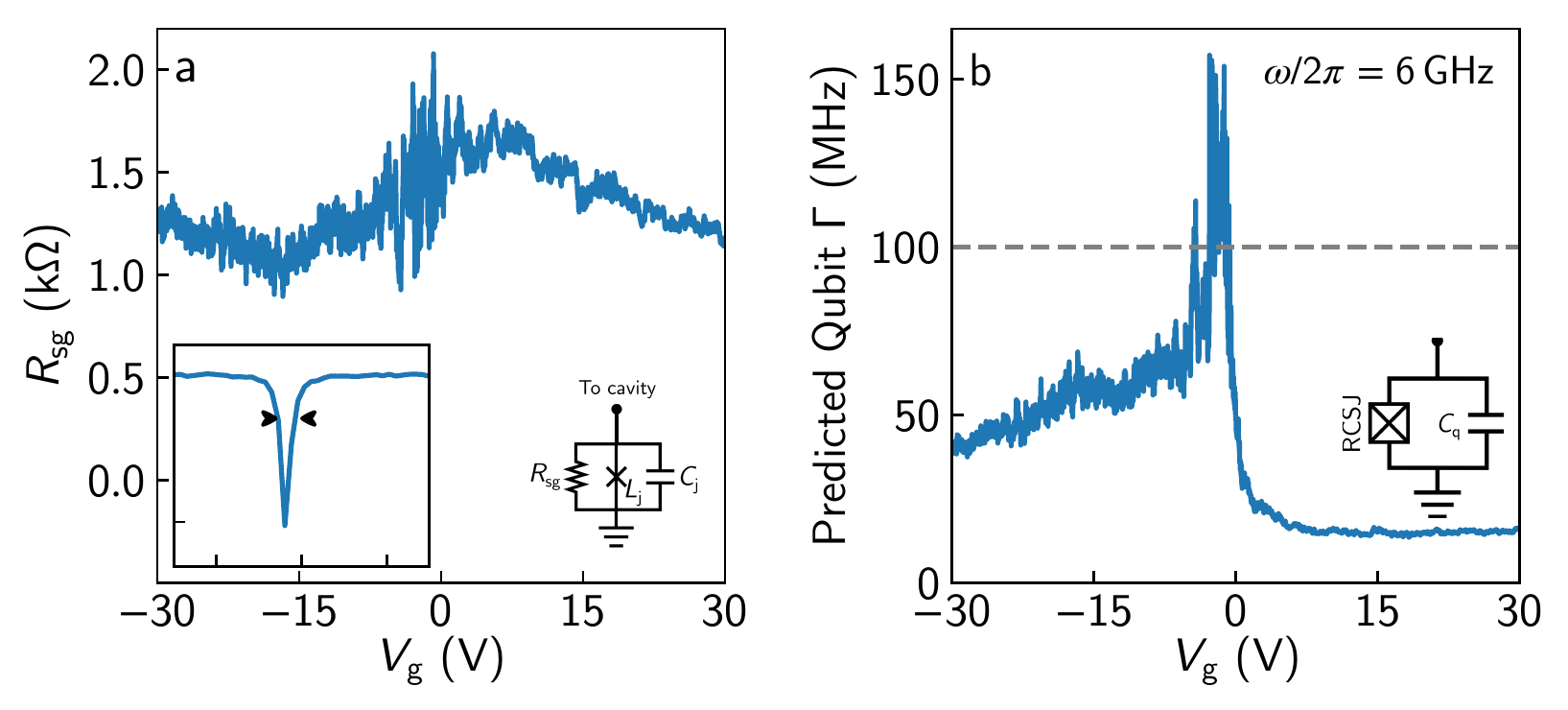}
	\caption[]{\textbf{Subgap resistance from microwave cavity measurements.} 
		\textbf{a,}
		Extracted sub-gap resistance at as a function of gate voltage.
		The values are calculated by calibrating the cavity properties and using the junction model shown connected to the transmission line cavity to fit the observed cavity response.
		Inset shows the cavity response at $V_\textrm{g}=\SI{30}{V}$.
		The horizontal and vertical axis divisions are \SI{10}{MHz} and \SI{10}{dB} respectively.
		\textbf{b,}
		Predicted linewidth for a graphene transmon qubit, obtained by taking the RCSJ parameters as a function of gate and adding a capacitance $C_\textrm{q}$ such that the final operating frequency remains $\omega/2\pi = \left(2\pi\sqrt{(L_\textrm{j}(C_\textrm{j}+C_\textrm{q}))}\right)^{-1} = \SI{6}{GHz}$.
		We assume the internal junction losses dominate the total linewidth.
		The horizontal line represents the anharmonicity of a typical SIS transmon $E_\textrm{c}/h=\SI{100}{MHz}$.
		In regions where the blue line falls under the dashed line, a gJJ transmon would be capable of operating as a qubit.
		The error bands for both panels can be found in Supplementary Fig. 5.
	}
	\label{fig:figure4}
\end{figure}

As shown in Fig. \ref{fig:figure4}a, the zero-bias sub-gap resistance is of the order of 1-\SI{2}{\kilo\Omega} and remains relatively flat on the range of applied gate voltages.
We find that the ratio $R_{\rm sg}/R_\textrm{n}$ has values around 10-40, depending on gate voltage, with higher values in the n-doped regime.
This ratio is often taken as figure of merit in SIS literature, as lower values of $R_{\rm sg}$ are detrimental to most applications since they imply higher leakage currents in DC and more dissipation in RF.

While $R_{\rm sg}$ of our device is lower than what would be implied by the coherence times in qubits based on low-critical-current oxide SIS junctions \cite{paik_observation_2011a}, the $R_{\rm sg}/R_\textrm{n}$ ratio is comparable to typical values from DC measurements of SIS devices with larger critical currents \cite{iosad_characterization_2002a,tolpygo_subgap_2013}.

The finite sub-gap resistance in superconductor-semiconductor devices is not fully understood, but is thought to originate from imperfect contact transparency, charge disorder and anti-proximity effects \cite{liu_phenomenology_2017a,bretheau_tunnelling_2017a}.
While state-of-the-art SNS devices based on epitaxial semiconductors only recently exhibited hard induced gaps \cite{chang_hard_2015,kjaergaard_quantized_2016}, there are to our knowledge no reports of this on graphene devices, suggesting an interesting direction for future research.
Another effect leading to finite sub-gap conductance is the size of our device, which is much larger and wider than usually employed junctions in microwave circuits.
Depending on the ratio of $\Delta$ to the effective round-trip time of sub-gap states across the junction, the Thouless energy $E_{\rm th}$, the sub-gap density of states can be non-negligible.

From previous reports\cite{rosdahl_andreev_2018}, and from simulations of our channel (see Supplementary Note 7 and Supplementary Fig. 12), it is expected that there are a number of low-lying sub-gap states that could limit the value of $R_{\rm sg}$.
This suggests that the losses could be reduced ($R_{\rm sg}$ increased) by moving towards the short junction regime in which the energies of these states are increased and hence a harder gap forms.
To maintain the same inductance $L_\textrm{j}$, the junction would also have to be made narrower to compensate for the higher critical currents associated with a shorter junction.
This would presumably further enhance $R_{\rm sg}$ since low-lying sub-gap states typically originate from states with high transverse momentum.
Given the fact that the geometry and aspect ratio of our junction is not at the limit of state-of-the-art fabrication capabilities, reducing the size is a promising step to reduce the losses in future gJJ based devices.

We finally analyse the potential performance of our device for circuit quantum electrodynamics (cQED) applications.
We consider the performance of a hypothetical transmon qubit \cite{koch_chargeinsensitive_2007} using the inductance of our gJJ operating at $\omega/2\pi=\SI{6}{GHz}$.
Assuming that the qubit losses are dominated by $R_{\rm sg}$, the quality factor of such a device is given by $R_{\rm sg}/(\omega L_\textrm{j})$ which in our case is of the order of a few hundred, a reasonable value considering further optimization steps can be taken.
In order to qualify as a qubit, the resonator linewidth should be smaller the transmon anharmonicity, given by the charging energy $E_\textrm{c}$.
In Fig. \ref{fig:figure4}b, we compare the predicted gJJ transmon linewidth $\Gamma$ with a typical value for the anhamonicity of SIS transmon qubits, $E_\textrm{c}/h = \SI{100}{MHz}$.
For a wide range of gate voltages, we find that the predicted linewidth is smaller than the anhamonicity, $\Gamma < E_\textrm{c}/h$, a promising sign for qubit applications of the technology.
We note, however, that the critical currents of this junction would be too high at large gate voltages (i.e. our Josephson inductances are too low), requiring a capacitor that would be too large to satisfy the condition $E_\textrm{c}/h \geq \SI{100}{MHz}$ and a resonant frequency of \SI{6}{GHz}.
To reduce the critical current (and increase the Josephson inductance), a narrower junction could be used, which could also increase the subgap resistance, further improving the performance.
A more in-depth discussion on this point is included in Supplementary Notes 3-5 and Supplementary Figs. 6-8.
We believe that implementing a graphene transmon qubit with good coherence times is feasible for future devices.
We also note that while the ballistic nature of the junction is not crucial for its operation in the microwave circuit, the lack of electronic scattering in the channel offers a nice platform to better understand the loss channels in comparison to highly disordered systems, with a potential to use this knowledge in the future to optimize devices. 

\section{Discussion}

In summary, we have measured a ballistic encapsulated graphene Josephson junction embedded in a galvanically accessible microwave cavity.
The application of an electrostatic gate voltage allows tuning of the junction critical current as well as the cavity resonance frequency through the Josephson inductance $L_\textrm{j}$.
While the DC response of the junction is broadly in line with previous work \cite{calado_ballistic_2015a,benshalom_quantum_2015,lee_ultimately_2015}, the RF measurement of the cavity-junction system provides additional information on $L_\textrm{j}$ and microwave losses in this type of junction.
A comparison of the DC and RF derived values of $L_\textrm{j}$ reveal deviations from sinusoidal current phase relations, including suggestions of features not previously observed, demonstrating that microwave probes can reveal new information about the junction physics. 
From the microwave losses of the resonance, we have extracted the junction sub-gap resistance and predicted that, with some optimization, it should be possible to make a coherent qubit based on a gJJ.
From the physics of the proximity junctions, we have suggested a route towards improving the coherence potentially towards the current state-of-the-art, enabling a new generation of gate-tunable quantum circuit technology.

\section*{Methods}
\subsection*{Fabrication of the microwave circuit}
We closely follow a recipe published earlier \cite{bosman_broadband_2015,singh_molybdenumrhenium_2014}.
In short, a \SI{50}{nm} film of MoRe is first sputtered onto a 2" sapphire wafer (\SI{430}{\micro m}, c-plane, SSP from \textit{University Wafers}).
The coplanar waveguide (CPW) resonator is defined using positive e-beam lithography and dry-etching with an $\mathrm{SF_6 + He}$ plasma.
We subsequently deposit \SI{60}{nm} of $\mathrm{Si_3N_4}$ for the shunt dielectric using PECVD and pattern this layer with a negative e-beam step and a $\mathrm{CHF_3 + O2}$ plasma.
The top plate of the shunts consists of a \SI{100}{nm} layer of MoRe which is deposited using positive e-beam lithography and lift-off.
An additional shunt capacitor, identical to the one on the main input, is built on the gate line.
This will filter RF noise on the gate line and suppress microwave losses through this lead.
Finally, we dice the wafer into $\SI{10}{mm}\times\SI{10}{mm}$ pieces, onto which the BN/G/BN stacks can be deposited.

\subsection*{Fabrication of the gJJ}
We exfoliate graphene and BN from thick crystals (HOPG from \textit{HQ Graphene} and BN from \textit{NIMS}\cite{taniguchi_synthesis_2007}) onto cleaned Si/$\mathrm{SiO_2}$ pieces using wafer adhesive tape.
After identifying suitable flakes with an optical microscope, we build a BN/G/BN heterostructure using a PPC/PDMS stamp on a glass slide \cite{pizzocchero_hot_2016,wang_onedimensional_2013}.
The assembled stack is then transferred onto the chip with the finished microwave cavity.
Using an etch-fill technique ($\mathrm{CHF_3 + O_2}$ plasma and NbTiN sputtering), we contact the center line of the CPW to the graphene flake on one side, and short the other side to the ground plane.
Clean interfaces between the NbTiN junction leads and the MoRe resonator body are ensured by maximizing the overlap area of the two materials and immediate sputtering of the contact metal after etch-exposing the graphene edge.
The resistance measured from the resonator center line to ground is therefore due entirely to the gJJ.
After shaping the device ($\mathrm{CHF_3 + O_2}$ plasma), we cover it with two layers of HSQ \cite{nanda_currentphase_2017} and add the top-gate with a final lift-off step.

\subsection*{Measurement setup}\label{sec:setup}
\noindent A sketch of the complete measurement setup is given in Supplementary Fig. 1.
The chip is glued and wire-bonded to a printed circuit board, that is in turn enclosed by a copper box for radiation shielding and subsequently mounted to the mK plate of our dry dilution refrigerator.
All measurements are performed at the base temperature of \SI{15}{mK}.
Using a bias-tee, we connect both the RF and DC lines to the signal port of the device while a voltage source is connected to the gate line.

We perform the microwave spectroscopy with a Vector Network Analyser (Keysight PNA N5221A).
The input line is attenuated by \SI{53}{dB} through the cryogenic stages, and \SI{30}{dB} room temperature attenuators.
Adding to these numbers an estimate for our cable and component losses results in a total attenuation on our input line of approximately \SI{92}{dB}.
The sample is excited with \SI{-30}{dBm}, so less than \SI{-122}{dBm} should arrive at the cavity.
This corresponds to an estimated intra-cavity photon number of at most 10-20 depending on operating frequency and linewidth (see Supplementary Fig. 9).
Test were run at $V_{\rm g} = 30$ V for powers down to \SI{-152}{dBm}, or approximately 0.02 photons, with negligible changes to the cavity line shape.
Other gate voltages are expected to have even lower photon populations for with the same setup due to the lower internal cavity Q-factor.
The reflected microwave signal is split off from the exciting tone via a directional coupler, a DC block, two isolators and a high-pass filter to reject any low-frequency noise coupling to the line.
The signal is furthermore amplified by a \SI{40}{dB} \textit{Low-Noise Factory} amplifier on the \SI{3}{K} plate, and two room-temperature \textit{Miteqs}, each about \SI{31}{dB}, leading to a total amplification of \SI{102}{dB}.
During all RF measurements, the bias current is set to zero.

The DC lines consist of looms with twelve twisted wire pairs, of which four single wires are used in the measurements presented here.
The lines are filtered with $\pi$-filters inside the in-house built measurement rack at room-temperature, and two-stage RC and copper-powder filters, thermally anchored to the mK plate.
To reduce the maximum possible current on the gate line, a \SI{100}{k\Omega} resistor is added at room-temperature.
For the DC measurements presented, we turn the output power of the VNA off and current-bias the gJJ, while measuring the voltage drop across the device with respect to a cold ground on the mK plate.

\subsection*{Data visualization}
To remove gate-voltage-independent features such as cable resonances, we subtracted the mean of each line for constant frequency with outlier rejection (\SI{40}{\percent} low, \SI{40}{\percent} high) from the original data, resulting in Fig. \ref{fig:figure2}(c).
All figures representing data are plotted using \textit{matplotlib} v2 \cite{hunter_matplotlib_2007}.

\section*{Data Availability}
All raw and processed data as well as supporting code for processing and figure generation is available in Zenodo with the identifiers \verb|doi:10.5281/zenodo.1296129|\cite{zenodo1} and \verb|10.5281/zenodo.1408933|\cite{zenodo2}.


\section*{Acknowledgements}
We acknowledge T\'omas \"Orn Rosdahl and Anton Akhmerov for discussions regarding theory and Srijit Goswami for discussions regarding fabrication.
This work was supported by the EU Graphene Flagship Program.
We additionally thank the Kavli Nanolab Delft for the cleanroom facilities.

\section*{Author contributions}
F.E.S and M.D.J fabricated the device and performed the measurements.
K.W. and T.T. supplied the BN bulk crystals.
F.E.S., M.D.J., and G.A.S. analysed and interpreted the data.
G.A.S. conceived the experiment and supervised the work.
F.E.S., M.D.J., and G.A.S. contributed to discussing and writing the manuscript.

\section*{Competing interests}
The authors declare no competing interests.

\section*{Figure legends}

\cleardoublepage

\makeatletter
\renewcommand*{\fnum@figure}{{\normalfont Supplementary Figure~\thefigure}}
\renewcommand*{\fnum@table}{{\normalfont Supplementary Table~\thetable}}
\makeatother
\renewcommand{\thesection}{Supplementary Note \arabic{section}}
\renewcommand{\bibsection}{\subsection*{SUPPLEMENTARY REFERENCES}}
\setcounter{section}{0}
\setcounter{figure}{0}

\noindent {\sf \centering \LARGE Supplementary Information for:

A ballistic graphene superconducting microwave circuit

\vspace{1em}

\large Schmidt and Jenkins et al.
}





\newpage

\begin{figure*}[]
    {\centering
    \includegraphics[width=.8\linewidth]{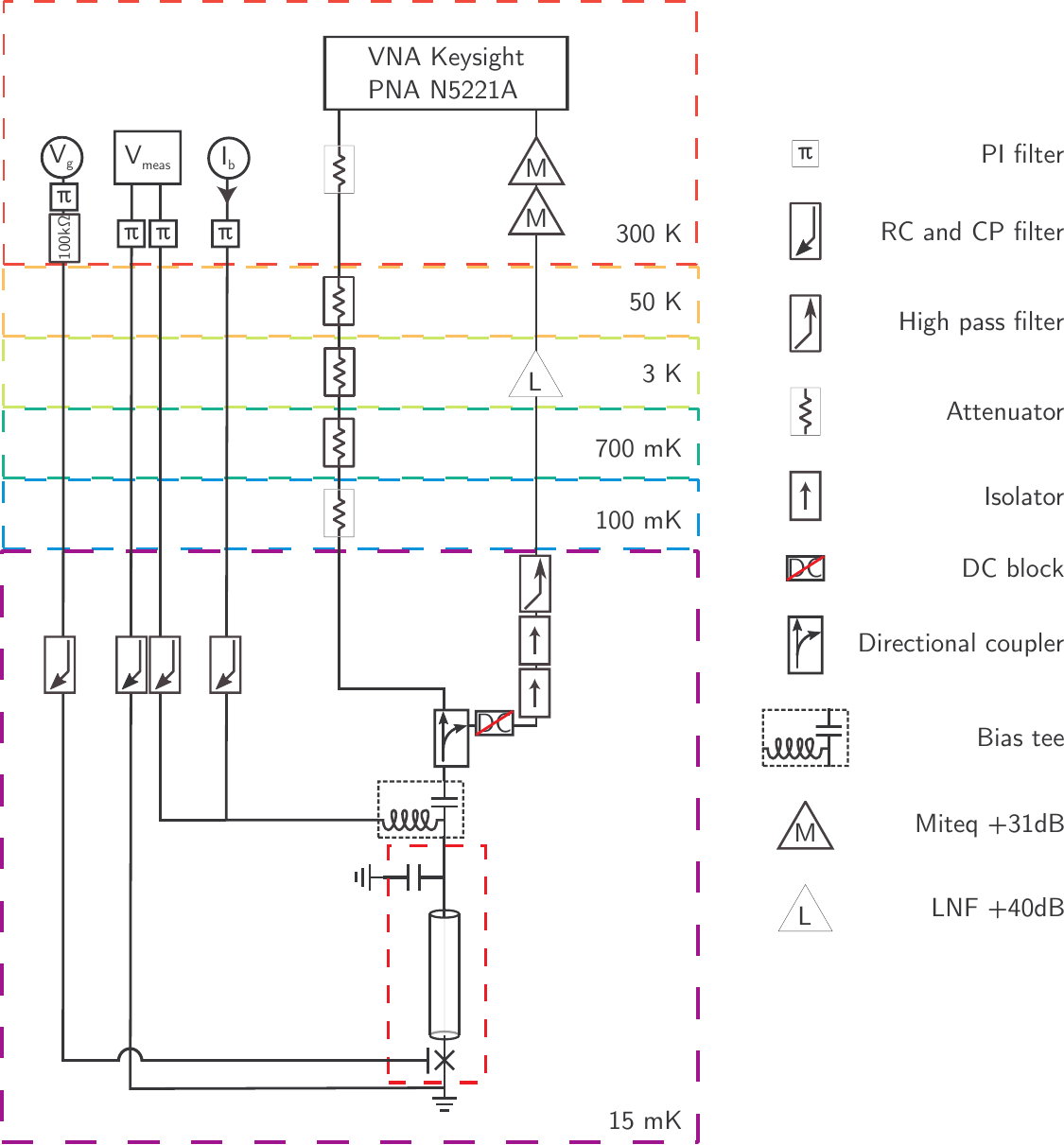}}
    \caption{{\bf Sketched measurement setup.}
        Dashed red box at the bottom marks device outline.
    }
    \label{fig:setup_full}
\end{figure*}

\begin{figure*}[]
    \centering
    \includegraphics[width=.7\linewidth]{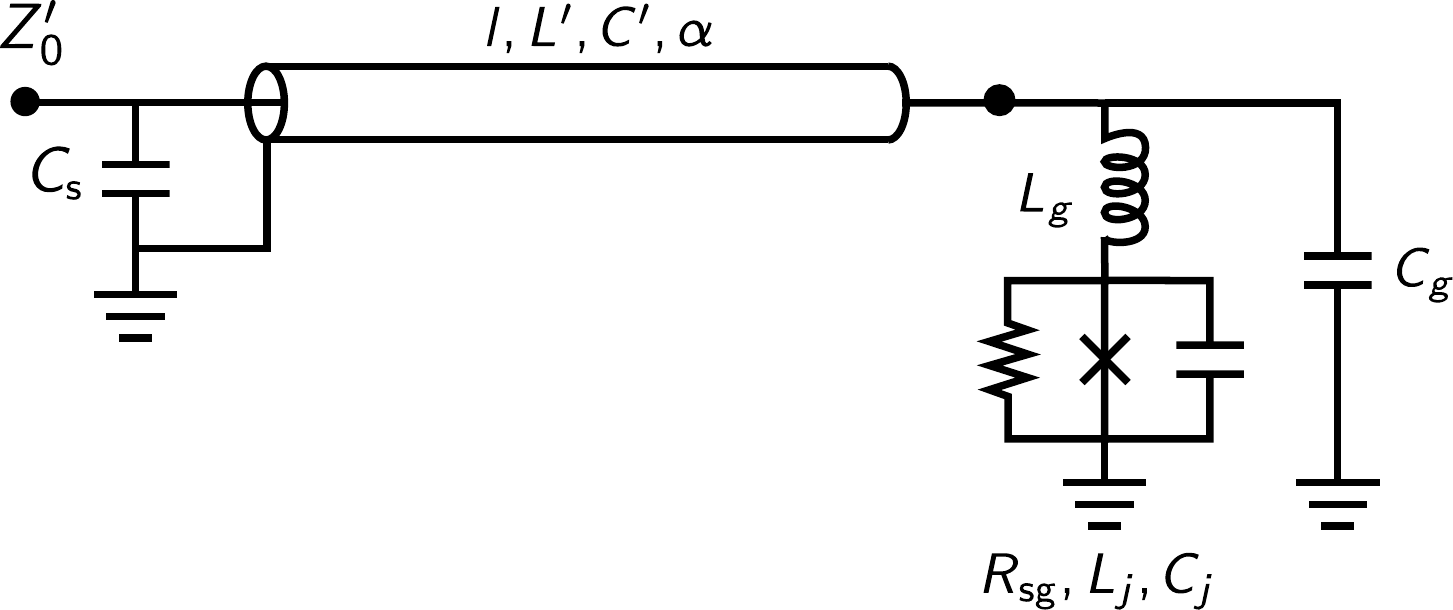}
    \caption[]{{\bf RF model for gJJ in cavity used for extraction of microwave parameters.}
        For the fitting procedure see \ref{sec:extraction}.
    }
    \label{fig:rfmodel}
\end{figure*}

\begin{figure*}[]
    \centering
    \includegraphics[width=\linewidth]{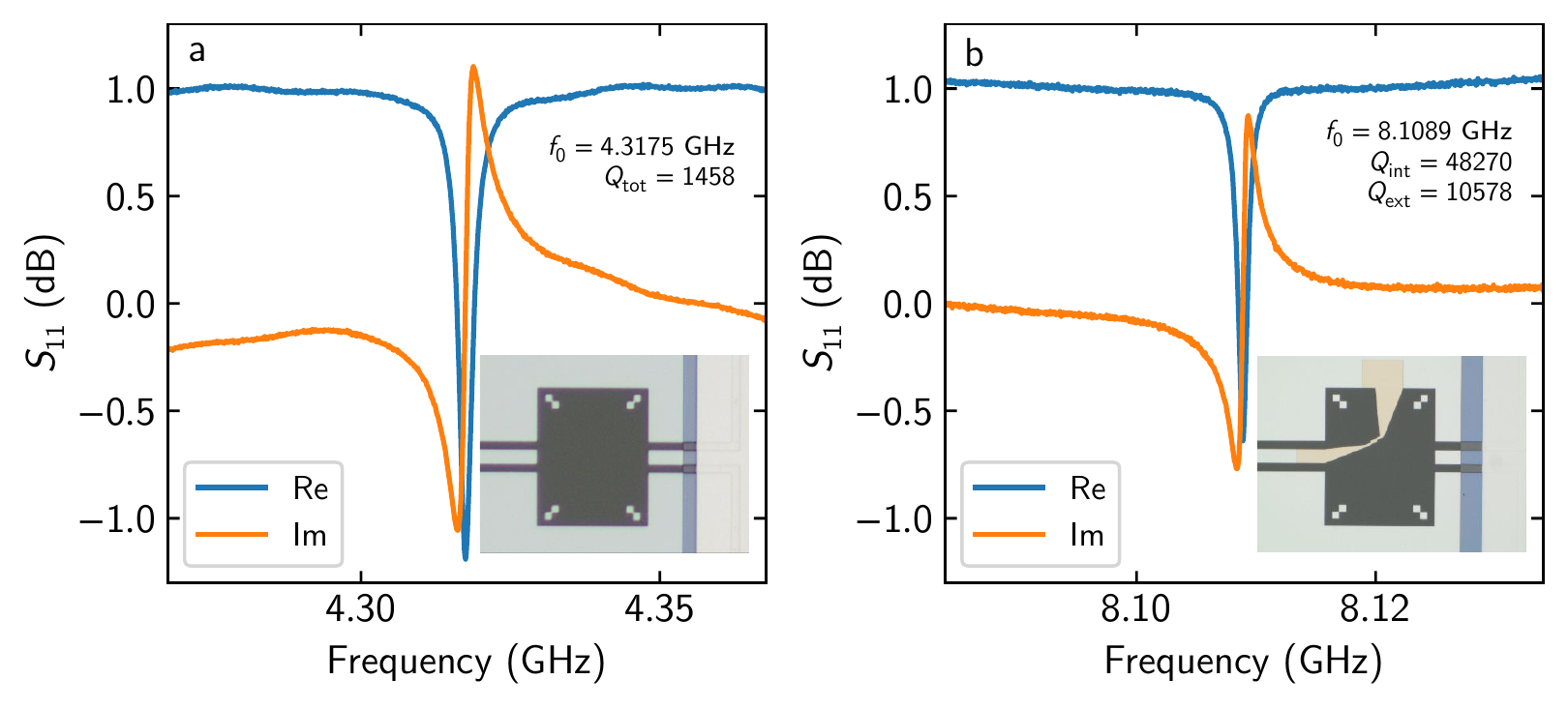}
    \caption[]{{\bf Reference samples for extraction of microwave parameters.}
        \textbf{a,} Open-ended cavity measurement of the real (imaginary) part of the reflection coefficient plotted in blue (orange).
        Inset: Optical micrograph of junction area of the measured device (open end).
        \textbf{b,} Shorted-cavity measurement with same lead geometry as the actual gJJ sample.
        Inset: Optical micrograph of junction area of the measured device (connected to ground).
    }
    \label{fig:calcavities}
\end{figure*}

\begin{figure*}[]
    \centering
    \includegraphics[width=0.5\linewidth]{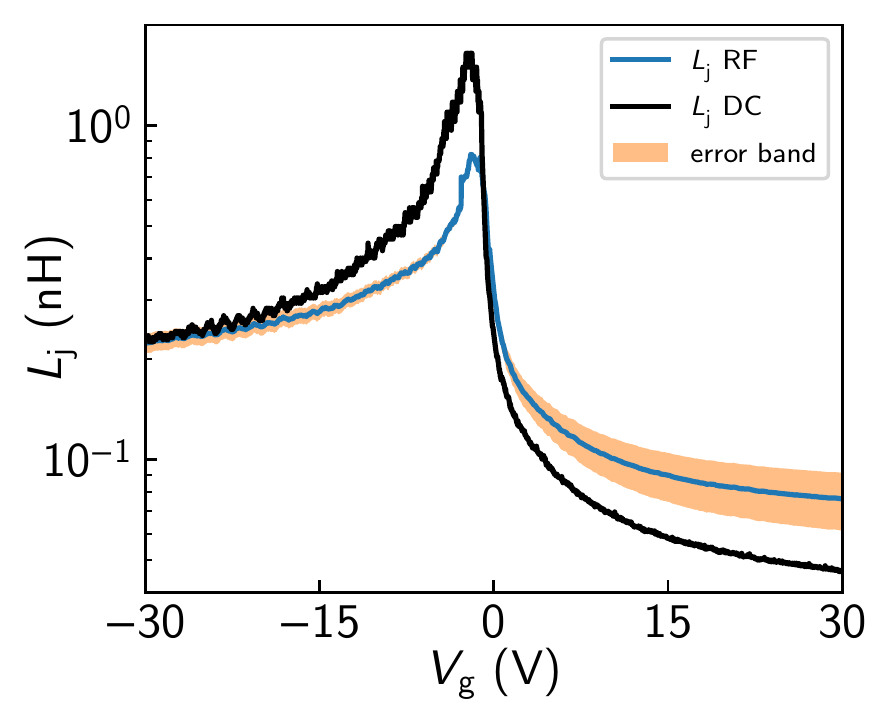}
    \caption[]{\textbf{Josephson inductance extracted from RF and DC measurements, including error bands.}
        We plot here the same quantities as in Figure 3 of the main text but include error bands corresponding to minimum and maximum values originating from uncertainties in the circuit.
        The scales are identical to the plots in the main text.
    }
    \label{fig:figure3_bands}
\end{figure*}

\begin{figure*}[]
    \centering
    \includegraphics[width=\linewidth]{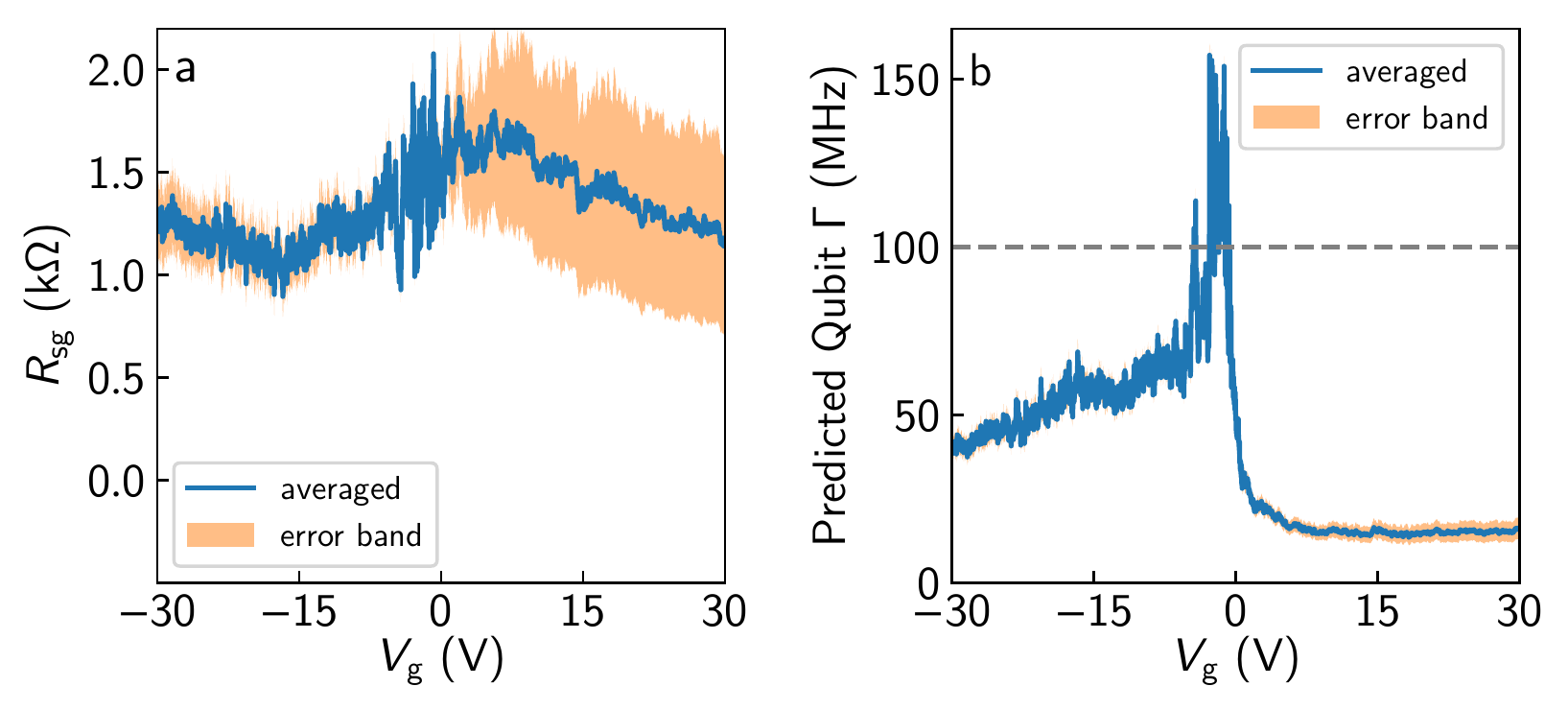}
    \caption[]{\textbf{Subgap resistance from microwave cavity measurements, including error bands.}
        We plot here the same quantities as in Figure 4 of the main text, but include error bands corresponding to minimum and maximum values originating from uncertainties in the circuit.
        The scales are identical to the plots in the main text.
        \textbf{a,} Subgap-resistance including error band.
        \textbf{b,} Corresponding linewidth of the hypothetical transmon with error band.
    }
    \label{fig:figure4_bands}
\end{figure*}

\begin{figure*}[]
    \centering
    \includegraphics[width=.7\linewidth]{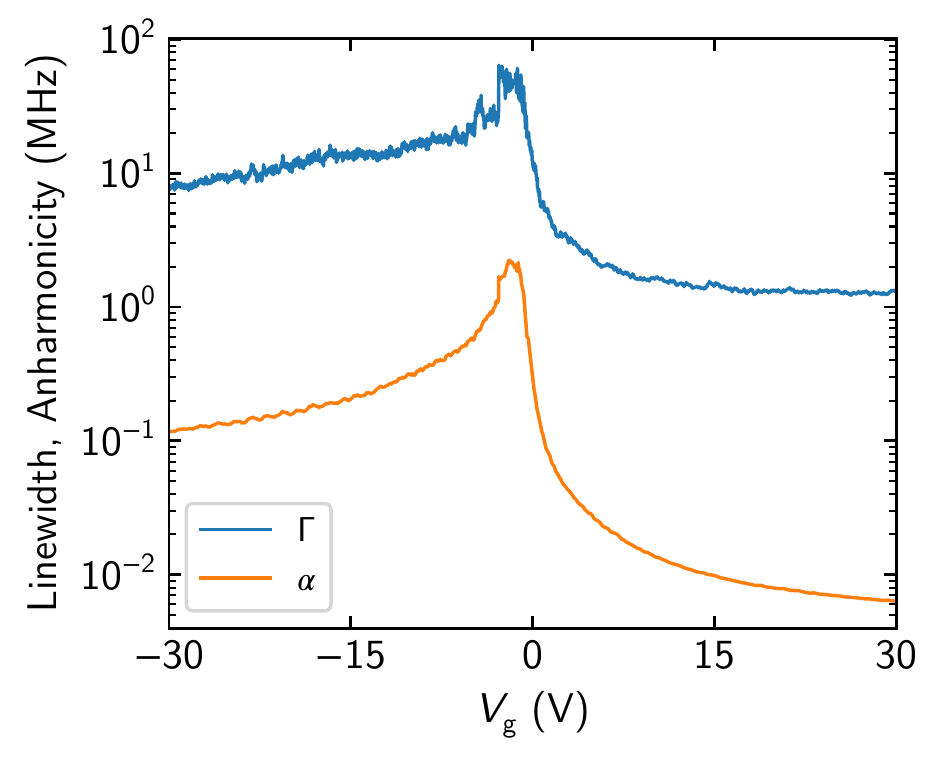}
    \caption{{\bf Anharmonicity and internal linewidth of current device, as described in \ref{sec:feasability}.}
    The calculated values of anharmonicity are always smaller than the measured linewidth meaning that this device cannot be considered a qubit in its current form.}
    \label{fig:anharm1}
\end{figure*}

\begin{figure*}[]
    \centering
    \includegraphics[width=.9\linewidth]{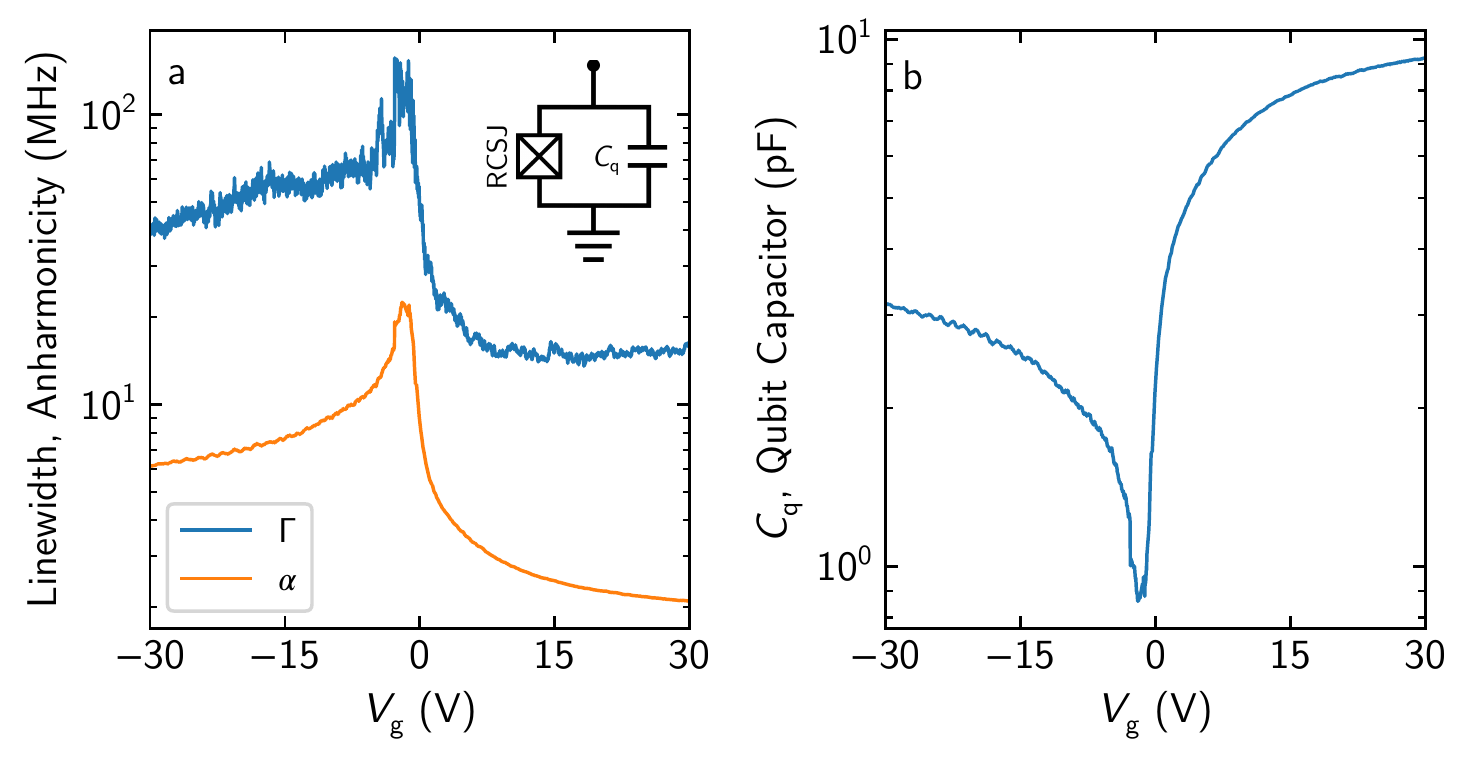}
    \caption{{\bf Anharmonicity and internal linewidth for design scenario A, as described in \ref{sec:scenA}.}
        \textbf{a,} We calculate the performance of the measured junction in a circuit shuch as the one shown in the inset.
        Setting the resonant frequency to $\omega_0 = 2\pi\cdot\SI{6}{GHz}$, we then calculate the anharmonicity and linewidth of this hypothetical device.
        Also in this case we find that calculated values of anharmonicity are always smaller than the linewidth.
        \textbf{b,} Required value of capacitance $C_{\rm q}$ to maintain a resonant frequency of $\omega_0 = 2\pi\cdot\SI{6}{GHz}$ as a function of $V_{\rm g}$}
    \label{fig:anharm2}
\end{figure*}

\begin{figure*}[]
    \centering
    \includegraphics[width=.9\linewidth]{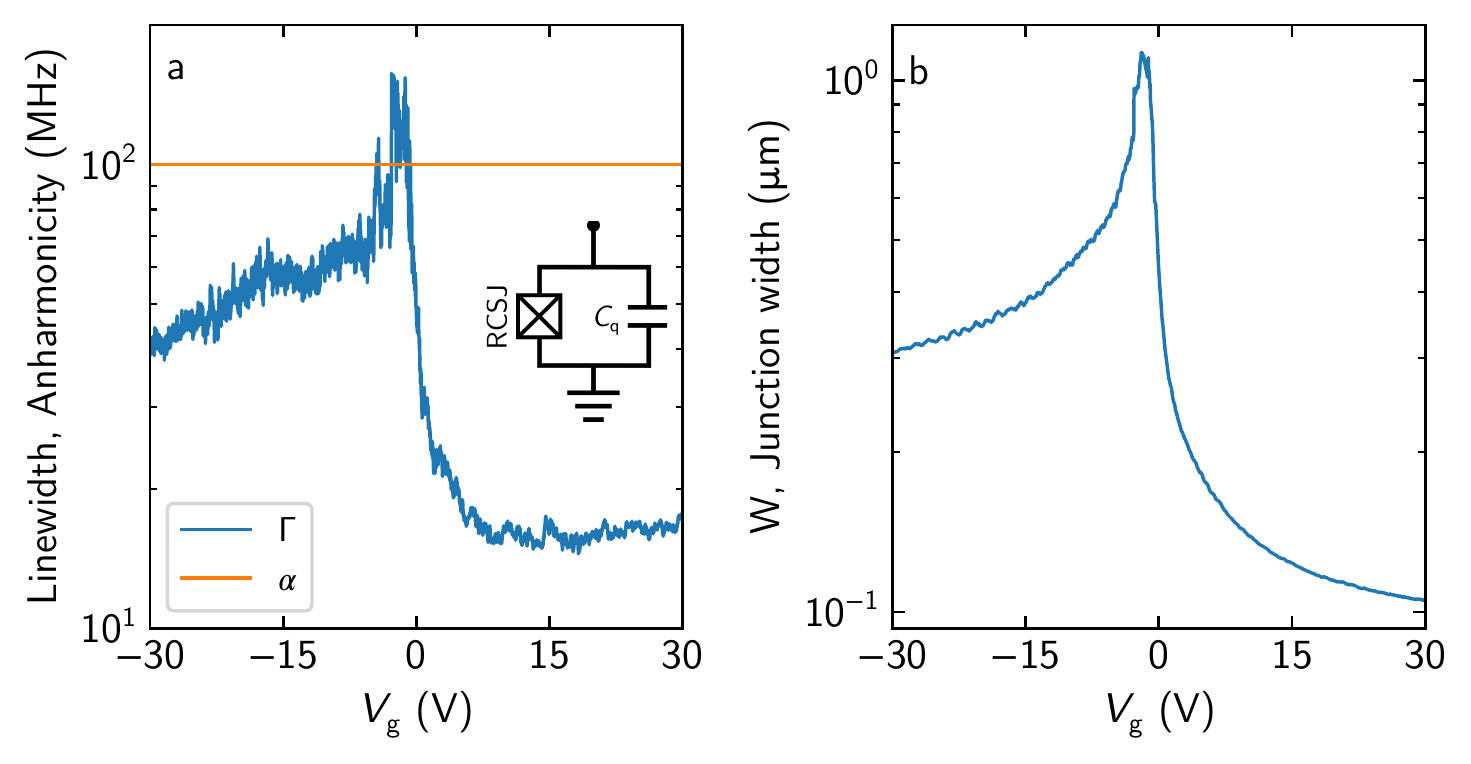}
    \caption{{\bf Anharmonicity and internal linewidth for design scenario B, as described in \ref{sec:scenB}.}
        \textbf{a,} We calculate the performance of a device whose capacitance and inductance are set by the requirement $\alpha = \SI{100}{MHz}$ and $\omega_0 = 2\pi\cdot\SI{6}{GHz}$.  This means scaling the junction width as a function of $V_{\rm g}$.  The expected linewidth $\Gamma$ is shown along with the designed anharmonicity.
        \textbf{b,} Required junction width to maintain a resonant frequency of $\omega_0 = 2\pi\cdot\SI{6}{GHz}$ and $\alpha = \SI{100}{MHz}$ as a function of $V_{\rm g}$.}
    \label{fig:anharm3}
\end{figure*}

\begin{figure*}[]
    \centering
    \includegraphics[width=0.6\linewidth]{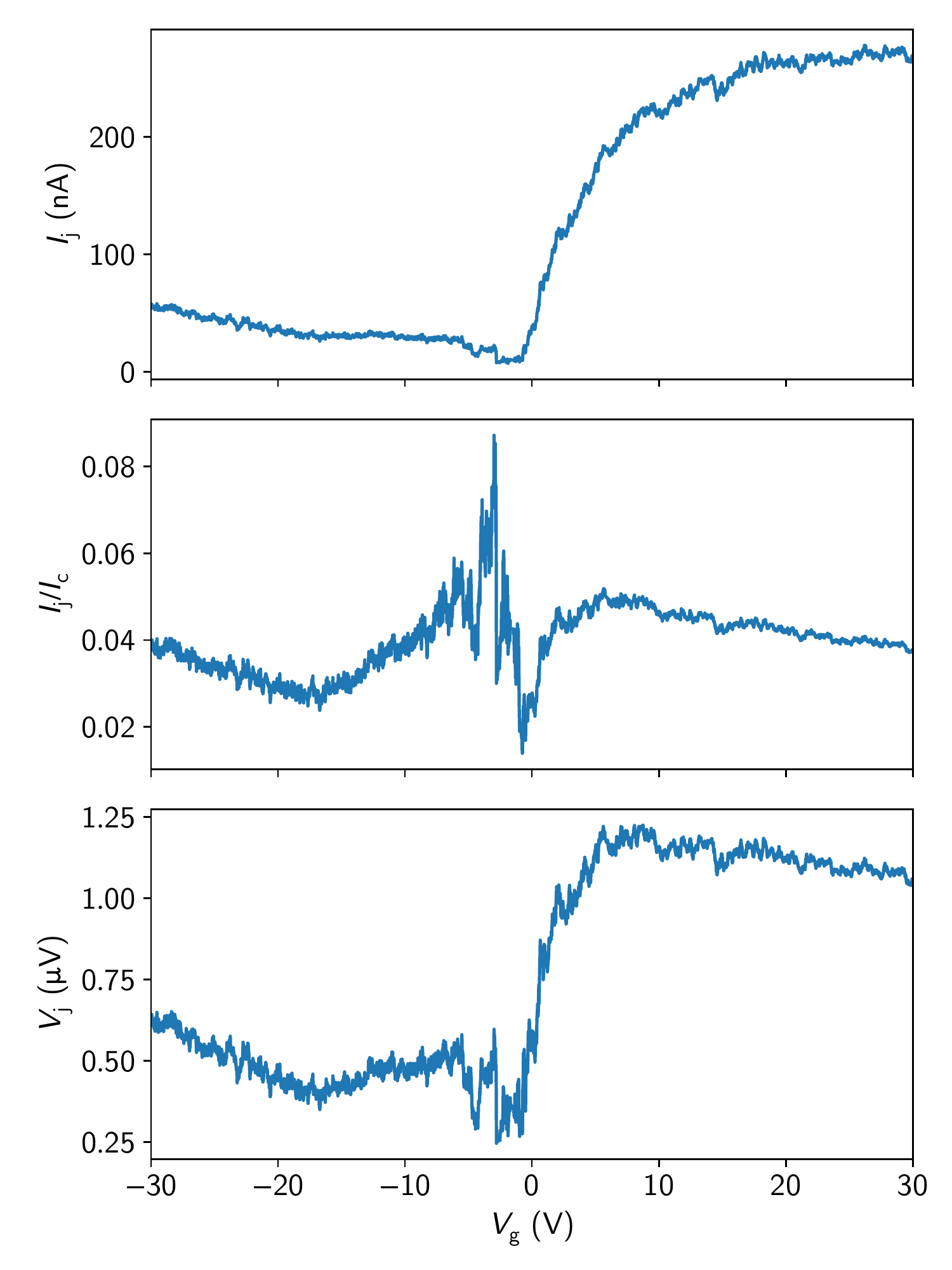}
    \caption{\textbf{Current and voltage amplitude at junction for measurement in Figure 2c of main text.}
        The input power at the device is estimated to be approximately \SI{-122}{dBm}.
        Currents are well below the measured critical current of the junction, even near the charge neutrality point.
        The average voltage across the junction induced by the microwave tone is lower than \SI{1}{\micro\volt}.
    }
    \label{fig:IjVj}
\end{figure*}

\begin{figure*}[]
    \centering
    \includegraphics[width=0.2\linewidth]{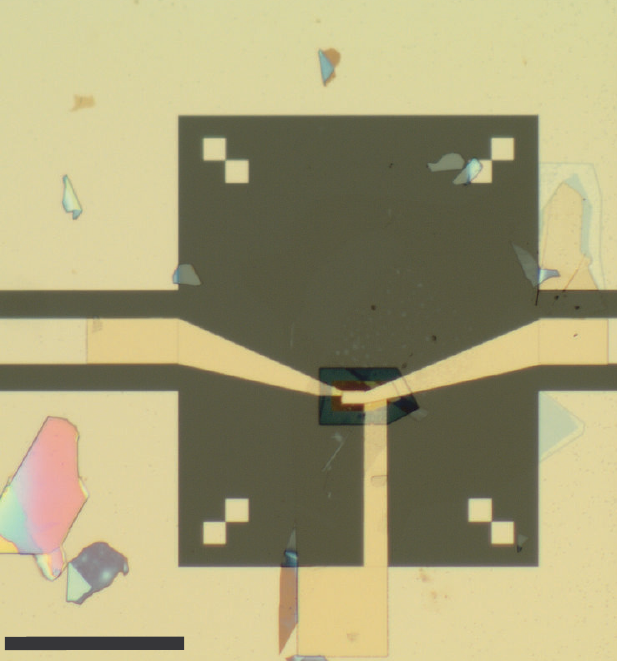}
    \caption{
        \textbf{Microscope image of second graphene superconducting junction.}
        The flakes around the device are hBN residues from the transfer process.
        Scale bar \SI{40}{\micro m}.
    }
    \label{fig:repeatdev_img}
\end{figure*}

\begin{figure*}[]
    \centering
    \includegraphics[width=\linewidth]{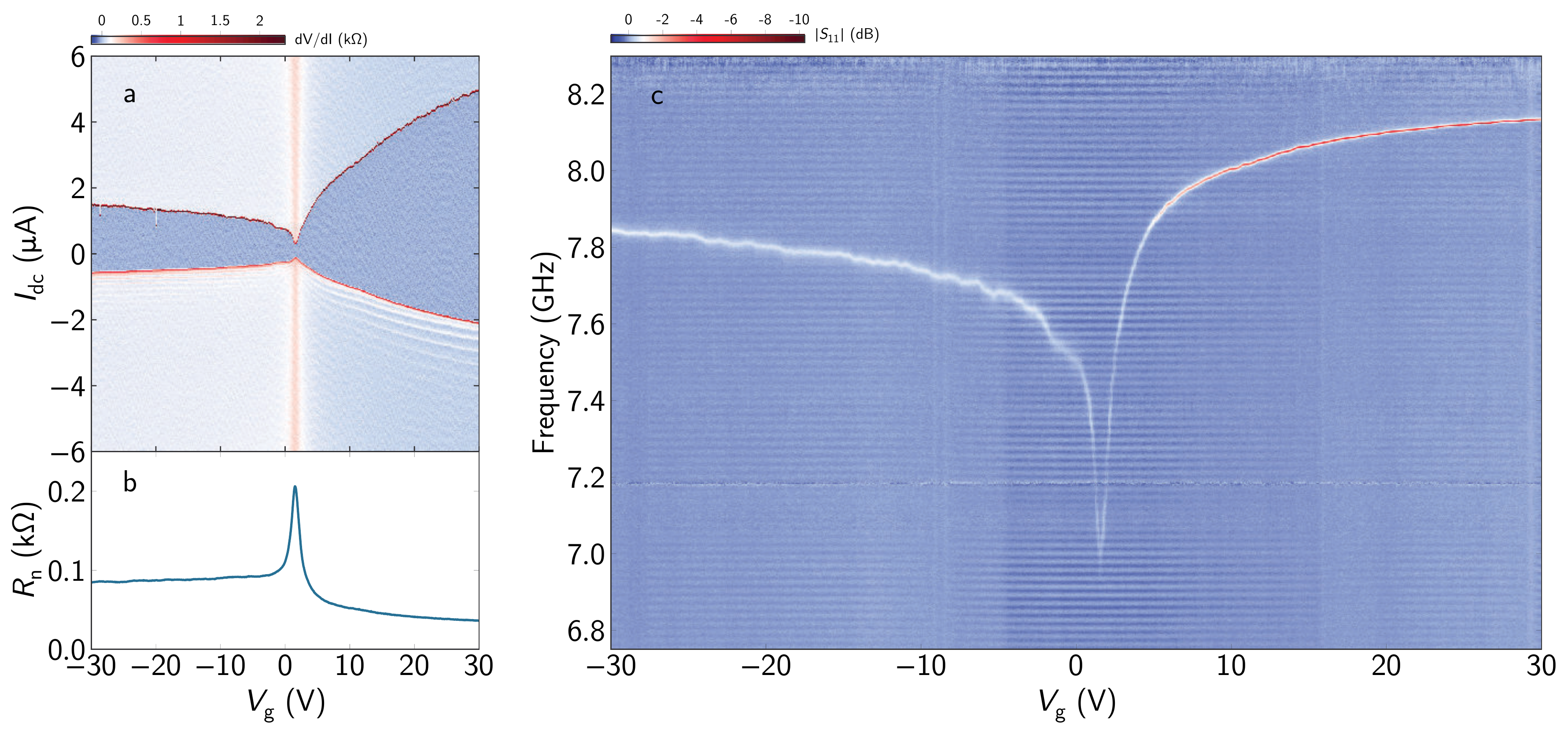}
    \caption{\textbf{Observation of the Josephson inductance of a second graphene superconducting junction. a,}
        Differential resistance across the gJJ (Supplementary Figure \ref{fig:repeatdev_img}) for a wide gate voltage range.
        Dark blue denotes area of zero resistance.
        \textbf{b,} Normal state resistance of the gJJ versus gate voltage.
        \textbf{c,} Microwave spectroscopy of the device in the superconducting state versus gate voltage, plotted as the amplitude of the reflection coefficient $\lvert S_{11} \rvert$ after background subtraction.
        Remarkably, its performance is broadly similar to the main text device (see Figure 2 of main text), despite having been stored at room temperature in a nitrogen box for ten months before measurement.
    }
    \label{fig:repeatdev}
\end{figure*}

\begin{figure*}[]
    \centering
    \includegraphics[width=\linewidth]{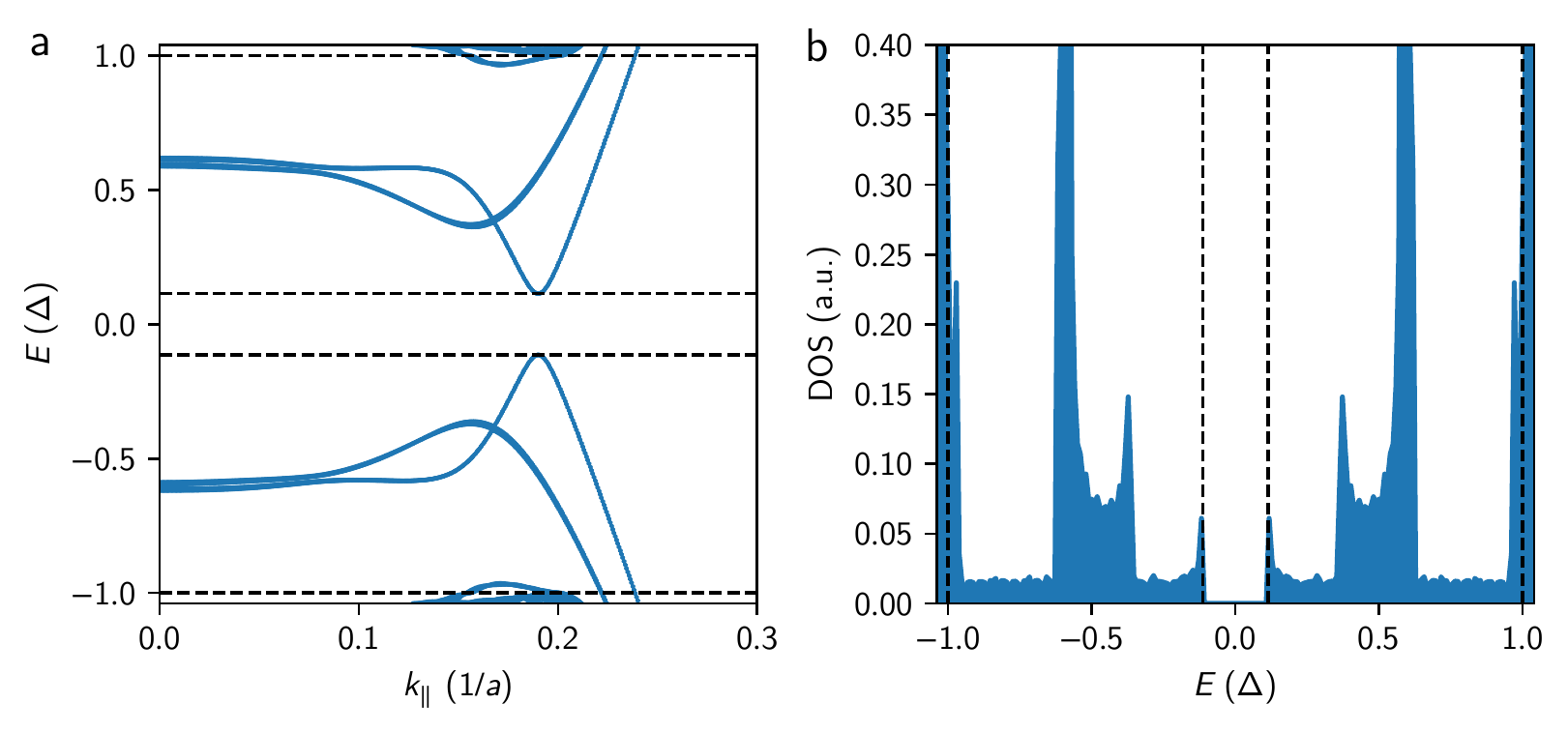}
    \caption{{\bf Dispersion and density of states of a gJJ, as described in \ref{sec:subgap}.}
        {\bf a}, Simulated subgap dispersion for a graphene junction in the intermediate regime, $\Delta/E_{\rm th}=1.542$ with $L_{\rm N}=60$ and infinite lateral extension.
        Energy is scaled with respect to $\Delta$, $k_\parallel$ in terms of momentum parallel to the SN-interface.
        {\bf b}, By binning the energy dispersion we obtain the density of states as a function of energy.
        Various subgap peaks originating from ABS with high transverse momentum occur, while a hard gap remains, as indicated by the dashed horizontal lines.
    }
    \label{fig:subgap_dos}
\end{figure*}

\begin{figure*}[]
    \centering
    \includegraphics[width=.6\linewidth]{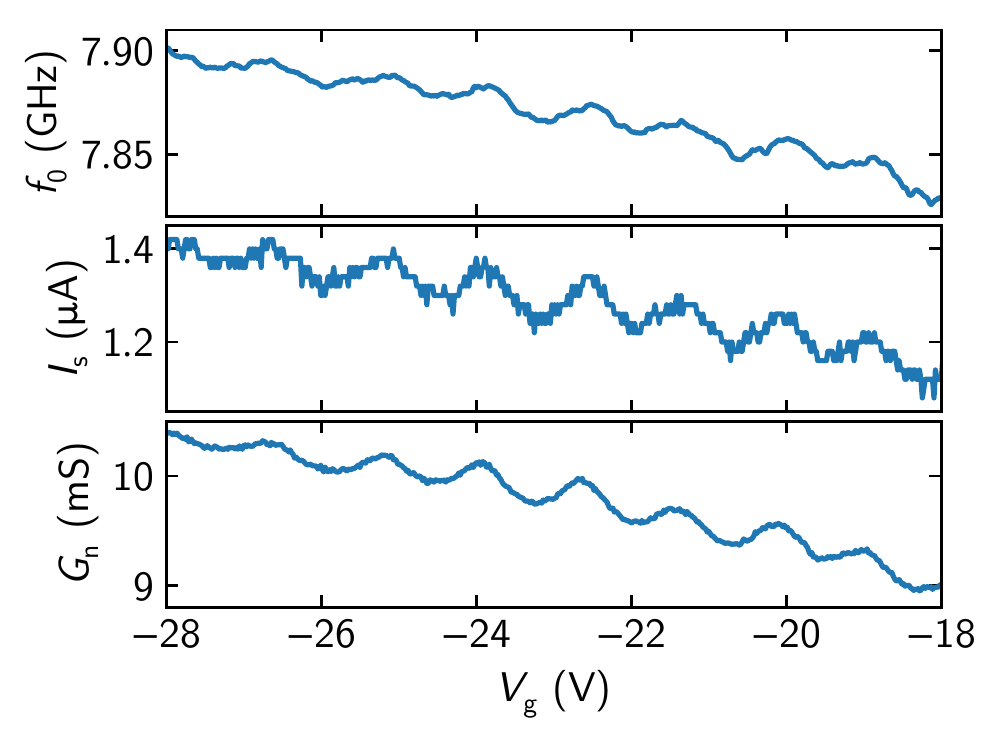}
    \caption{\textbf{Correlating oscillations in DC and RF measurements.}
        We observe reproducible and matching oscillations in-phase oscillations of resonance frequency, critical current and normal state conductance in the npn-regime.
        We attribute these to interfering electron waves partially reflected from the SN interfaces at the graphene-superconductor contacts:
        Since NbTiN slightly n-dopes the contact region (hence the asymmetry in $R_{\rm n}$ as a function of gate voltage), pn-junctions form at the interface once the graphene is driven into the p-doped regime by the gate voltage.
        In the case of ballistic transport across the graphene sheet, the different charge carrier trajectories interfere with each other.
        Varying the gate voltage leads to a change in Fermi wavelength and hence an alternation of constructive and destructive interference, resulting in reduced and suppressed conductance, supercurrent, or inductance.
        This is akin to Fabry-P\'erot oscillations of light waves in free space, bound by two mirrors.
        The observation of these Fabry-P\'erot oscillations in graphene-based systems is uniformly taken as evidence of ballistic transport \cite{liang_fabry_2001,miao_phasecoherent_2007,young_quantum_2009,cho_massless_2011,wu_quantum_2012,campos_quantum_2012,rickhaus_ballistic_2013,benshalom_quantum_2015,calado_ballistic_2015a,amet_supercurrent_2016a,borzenets_ballistic_2016a,allen_observation_2017,zhu_supercurrent_2018}.
        We therefore conclude that our device is also in the ballistic regime.
        We analyse these oscillations in Supplementary Figure \ref{fig:fabry-perot}.
    } 
    \label{fig:quantum_osc}
\end{figure*}

\begin{figure*}[]
    \centering
    \includegraphics[width=\linewidth]{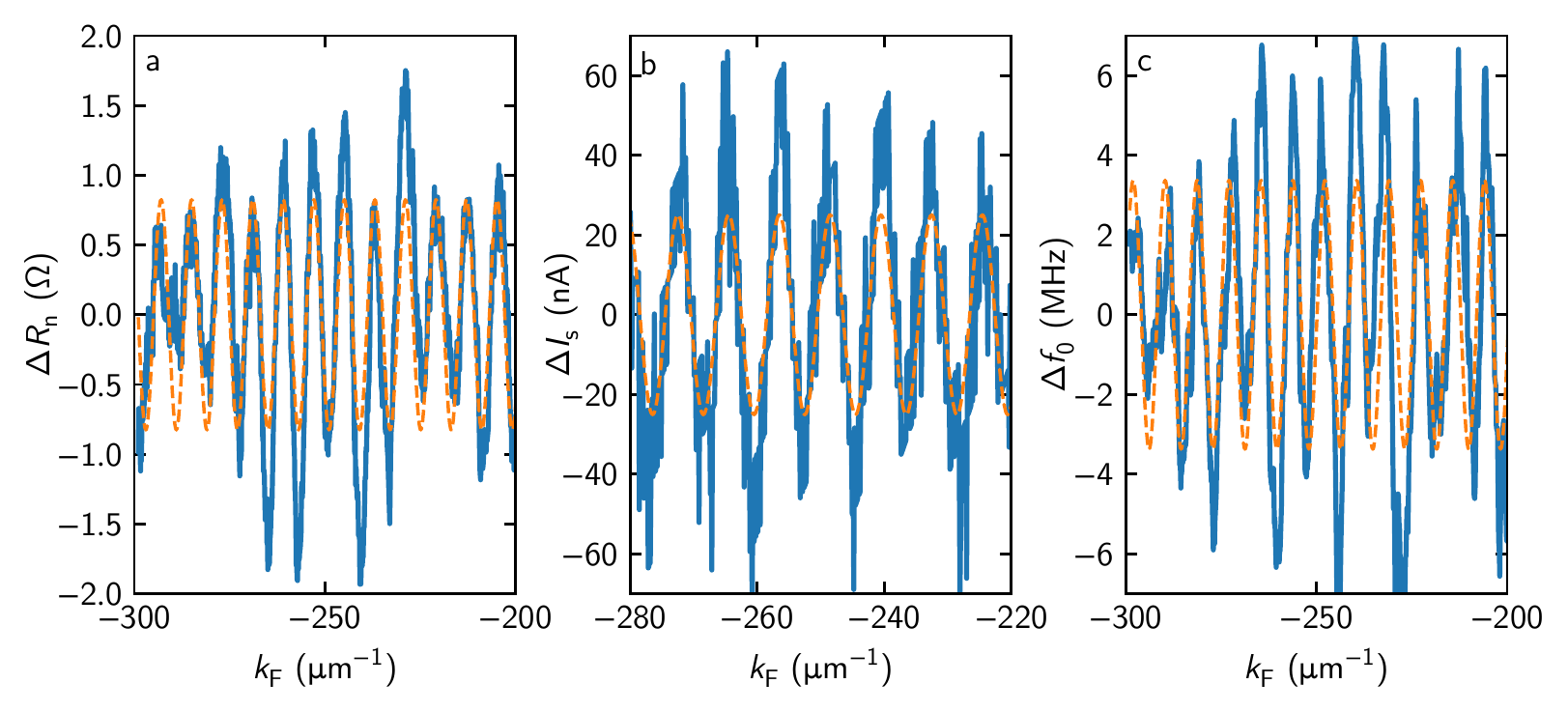}
    \caption{{\bf Fabry-P\'erot oscillations in ballistic gJJ.}
        We observe FP oscillations in \textbf{(a)} $R_{\rm n}$, \textbf{(b)} $I_{\rm c}$ and \textbf{(c)} $f_0$.
        We can extract the length of the resonant cavity by fitting our oscillating signal with a sine, according to the resonance condition $2L_{\rm c} = m\lambda_{\rm F}, m\in\mathbb{N} \rightarrow 2L_{\rm c} k_{\rm F} = 2\pi m$.
        After subtracting a slowly varying background with a third-order polynomial \cite{calado_ballistic_2015a}, the fits for $R_{\rm n}$, $I_{\rm c}$ and $f_0$ (orange lines) independently yield $L_{\rm c}\approx \SI{390}{nm}$.
        This suggests a contact interface barrier of no more than \SI{55}{nm} on each side.
        We can thus take $L_{\rm c}$ as a lower bound for the free momentum scattering and the phase coherence lengths, i.e. $l_{\rm mfp},\xi>L_{\rm c}$.        }
    \label{fig:fabry-perot}
\end{figure*}

\begin{figure*}[]
    \centering
    \includegraphics[width=\linewidth]{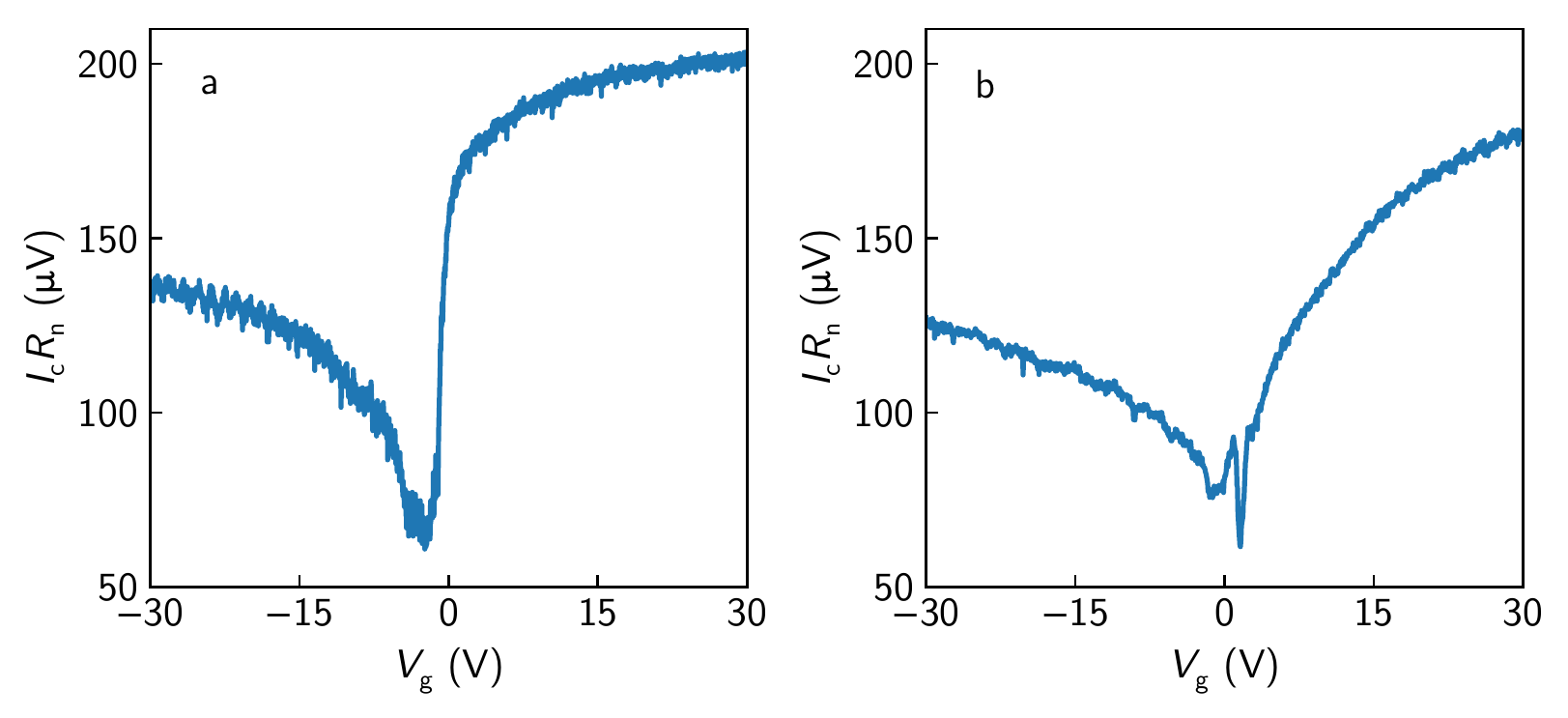}
    \caption{{\bf $I_{\rm c} R_{\rm n}$ product of gJJ devices.}
        The $I_{\rm c} R_{\rm n}$ product in Josepshon junctions is directly proportional to the gap voltage \cite{tinkham_introduction_1996}, with $I_{\rm c} R_{\rm n}\geq2.08\Delta/e$ in the case of ballistic graphene junctions \cite{titov_josephson_2006,cuevas_subharmonic_2006}.
        \textbf{a,} In our main device, this quantity saturates at approximately \SI{200}{\micro V} for high n-doping, drops to \SI{50}{\micro V} around CNP, and reaches up to \SI{130}{\micro V} for high p-doping.
        We take the small dependence on gate voltage in high doping regime as further indication of ballistic transport \cite{mizuno_ballisticlike_2013,zhu_supercurrent_2018}.
        Taking the bulk gap of the leads to be $\Delta=1.764 k_{\rm B} T_{\rm c} = \SI{2}{meV}$, our maximum $I_{\rm c} R_{\rm n}=0.1\Delta$ which is much lower than the theoretically expected value.
        We attribute this to reduced contact transparency and our junction being in the long regime, where the Thouless energy $E_{\rm th}=hv_{\rm F}/L < \Delta $ is the dominant energy scale, limiting $I_{\rm c} R_{\rm n}$ \cite{dubos_josephson_2001}.
        Our observation matches that of various other groups \cite{mizuno_ballisticlike_2013,benshalom_quantum_2015,borzenets_ballistic_2016a,zhu_supercurrent_2018}.
        \textbf{b,} In contrast, the additional device lacks the saturating behaviour, and exhibits a lower $I_{\rm c}R_{\rm n}$ product.
        This, in addition to the absence of FP oscillations, leads us to conclude that the latter device is non-ballistic, possibly due to a slightly longer normal region, or residual dirt (such as bubbles) in the graphene channel.
        }
    \label{fig:icrn}
\end{figure*}

\begin{figure*}[]
    \centering
    \includegraphics[width=.7\linewidth]{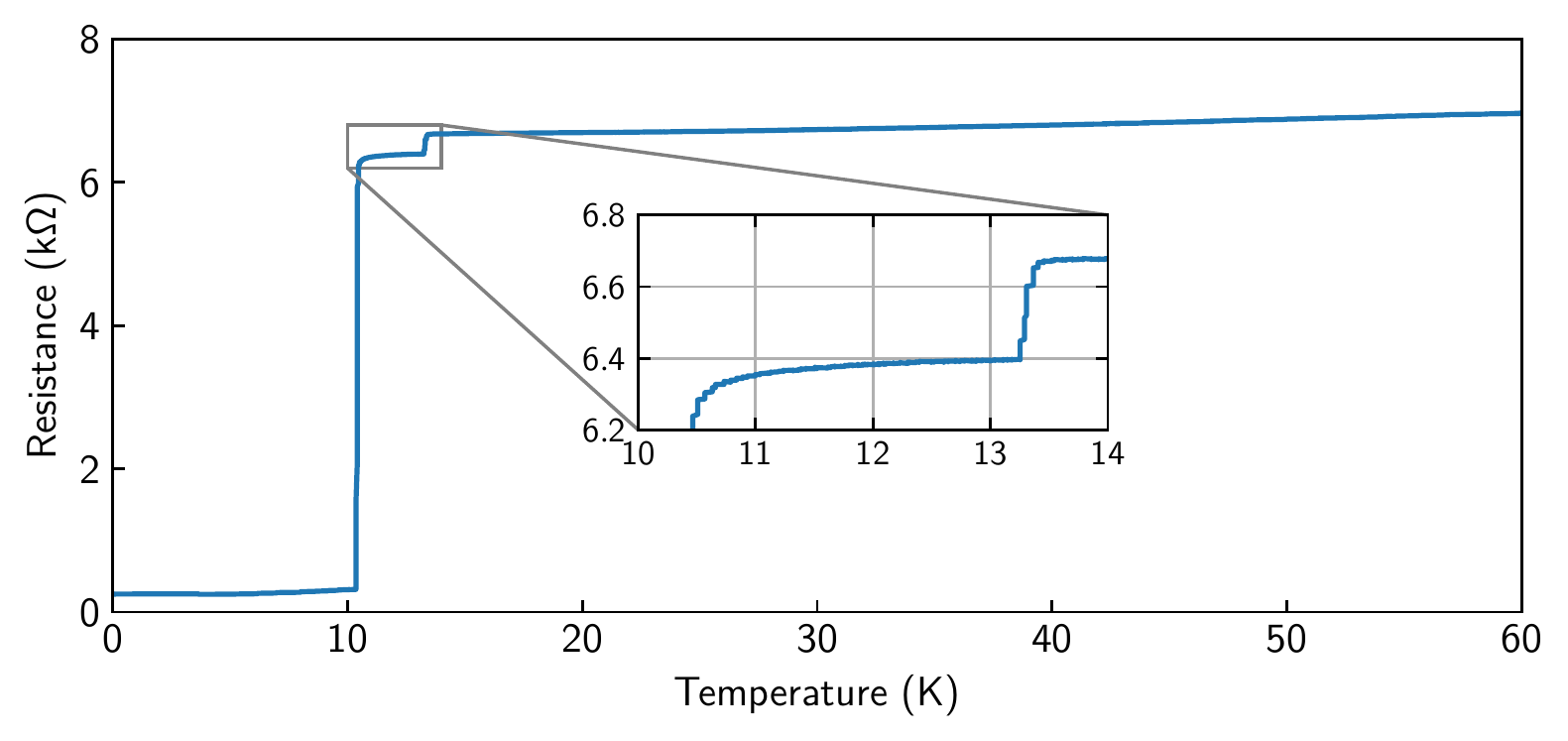}
    \caption{{\bf Critical temperature of MoRe and NbTiN.}
        Resistance versus temperature of the gJJ sample, measured during the initial cooldown, for a current bias of \SI{1}{\micro A} without any gate voltage applied.
        The two jumps at \SI{10.5}{K} and \SI{13.2}{K} correspond to the critical temperature of MoRe and NbTiN, respectively.
        Below $T_\mathrm{c,MoRe}$, we measure a residual resistance of \SI{250}{\Omega}, which corresponds to the graphene sheet resistance for $V_{\rm g}=\SI{0}{V}$.}
    \label{fig:critical_temp}
\end{figure*}


\begin{table*}
    \begin{center}
        \begin{tabular}{|c|c|}
            \hline 
            $l$ (TL length) & \SI{6119}{\micro\meter} \\ 
            \hline 
            $C'$ (Capacitance per unit length) & \SI{0.14848}{nF/m} \\ 
            \hline 
            $L'$ (Total inductance per unit length) & \SI{0.619838}{\micro\henry/m} \\ 
            \hline
            $C_{\rm s}$ (Shunt coupler capacitance) & $\sim\SI{27}{pF}$ \\ 
            \hline 
            $Z_0$ (TL Characteristic impedance) & \SI{64.611}{\ohm} \\
            \hline 
            $Z'_0$ (Reference impedance) & \SI{50}{\ohm} \\
            \hline 
            $v_{\rm ph}$ (Phase velocity in TL) & $\SI{1.04238e8}{m/s} = 0.3477\,c$ \\ 
            \hline
            $L'_{\rm g} = \frac{\mu_0}{4}\frac{K({k'_0}^2)}{K({k_0}^2)}$ (Geometric inductance per unit length) & \SI{0.4277}{\micro\henry/m} \\ 
            \hline
            $L'_{\rm k}$ (Kinetic inductance per unit length) & \SI{0.1922}{\micro\henry/m} \\ 
            \hline
            $L'_{\rm k}/L'$ (Kinetic inductance fraction) & \num{0.31} \\ 
            \hline
            $L_{\rm g}$ (Geometric inductance of junction leads) & \SIrange[range-phrase=--]{70}{100}{pH} \\ 
            \hline
            $C_{\rm g}$ (Geometric capacitance of junction leads) & \SI{4.7}{fF} \\ 
            \hline
            $C_{\rm j}$ (gJJ capacitance) & \SI{2}{fF} \\ 
            \hline
            $\alpha$ (Attenuation at \SI{8.1089}{GHz}) & \SI{0.006073}{\per\metre} \\ 
            \hline
        \end{tabular}
        \caption{\bf Transmission line, coupler and junction parameters with kinetic inductance correction included, as described in \ref{sec:extraction}.}
        \label{tab:tlpars}
    \end{center}
\end{table*}

\clearpage

\section{Fitting routine for extracting the resonance frequency}\label{sec:fitting}
\noindent The microwave response function of a capacitively shunted resonator in reflection geometry is given by \cite{pozar_microwave_2012}
\begin{eqnarray}
\Gamma(\omega) = \frac{\kappa_\mathrm{ext}-\kappa_\mathrm{int}-2i\Delta\omega}{\kappa_\mathrm{ext}+\kappa_\mathrm{int}+2i\Delta\omega},
\end{eqnarray}
where $\kappa_\mathrm{ext,int}=\omega_0/Q_\mathrm{ext,int}$ are the internal and external loss rates and $Q_\mathrm{ext,int}$ are the respective quality factors.
$\Delta\omega=\omega-\omega_0$ is the frequency detuning from the resonance frequency $\omega_0$.

The measured reflection coefficient must also include the effect of the connecting wires and devices between the network analyser and the device under test.
The reflection coefficient is accordingly modified to incorporate this background:
\begin{eqnarray}
S_{11} = B(\omega)\left(-1 + \frac{2\kappa_\mathrm{ext}e^{i\theta}}{\kappa_\mathrm{ext}+\kappa_\mathrm{int}+2i\Delta\omega}\right)
\end{eqnarray}
The complex background $B(\omega)$ has the form:
\begin{equation}
B(\omega) = (a+b\omega+c\omega^2)e^{i(a'+b'\omega)},
\end{equation}
where $a,b,c,a',b'$ are real parameters.  We use this function to fit the measurement data and extract $\omega_0$ and $\kappa_\mathrm{ext,int}$.

\section{Extraction of parameters from microwave measurements}\label{sec:extraction}
\noindent The schematic for the gJJ and cavity model can be seen in Supplementary Figure \ref{fig:rfmodel}.
A segment of a coplanar waveguide forms a cavity coupled on one side to an input line through a shunt capacitor.
The far end of the transmission line (TL) segment has the gJJ modelled using an RCSJ model with an extra inductance and capacitance associated to the junction lead wires.

The parameters needed to characterize the system are described below, listed in Supplementary Table \ref{tab:tlpars} and labelled in Supplementary Figure \ref{fig:rfmodel}:
\begin{itemize}
    \item The transmission line (TL) segment has a length $l$ as well as a capacitance per unit length $C'$ and inductance per unit length $L'$.
    TL losses are characterized by the attenuation parameter $\alpha$.
    It is worth noting that $L' = L'_{\rm g} + L'_{\rm k}$ includes a geometric contribution, $L'_{\rm g}$, and kinetic inductance contribution\cite{vanduzer_principles_1999}, $L'_{\rm k}$.
    \item The effective value of the shunt capacitance $C_{\rm s}$.
    Since $C_{\rm s}$ parametrizes the external cavity coupling, this includes contributions from both the shunt capacitor and the external circuit.
    The different connectors, wires, and other microwave components introduce impedance mismatches and cable resonances in the input/output lines, changing the external coupling.
    We use $C_{\rm s}$ to reabsorb most of these effects, hence making it frequency dependent.
    \item The characteristic impedance of the input line $Z'_0$ taken as \SI{50}{\ohm}, i.e., the VNA reference impedance.
    \item The gJJ is characterized by a junction inductance $L_{\rm j}$, a junction capacitance $C_{\rm j}$ and subgap resistance $R_{\rm sg}$.
    \item The junction leads also add a series inductance $L_{\rm g}$ and a shunt capacitance $C_{\rm g}$.
\end{itemize}

With these inputs, the reflection response of the circuit can be calculated analytically and compared to the measured data.
However, most of these parameters need to be calibrated and calculated first in order to deduce the junction parameters from the measurements.
The different parameters and calibrations are set as follows:
\begin{itemize}
    \item The cavity length is set by the design geometry of the cavity $l=\SI{6119}{\micro\meter}$ and verified through microscope inspection.
    \item To determine the cavity $L'$ and $C'$ as well as the internal losses (related to $\alpha$), several cavity measurements from the same batch as the final device were used.
    From fitting the fundamental mode resonances of these calibration samples we extracted values for $L'$, $C'$, $\alpha$ that we use for the final device.
    The samples used were:
    \begin{itemize}
        \item A cavity with no junction at the end (Supplementary Figure \ref{fig:calcavities}a).
        This means that the fundamental mode frequency is approximately half that of the final device ($\lambda/4$ vs $\lambda/2$ boundary conditions).
        From this measurement and the physical geometry of the cavity, we deduce values for $C'$, $L'$.
        \item A cavity with a short at the end with the same shape as the final junction leads (Supplementary Figure \ref{fig:calcavities}b).
        This cavity was used to calibrate the loss parameter $\alpha$ associated to resistive and dielectric losses of the transmission line cavity.
        In principle, the losses are frequency dependent with higher losses at higher frequencies.
        Since this loss rate was obtained at the high end of the frequency range and is used for all our frequencies, the extracted loss rates are expected to overestimate the actual losses.
    \end{itemize}
    \item The leads series inductance $L_{\rm g}$ and shunt capacitance $C_{\rm g}$ as well as the junction capacitance $C_{\rm j}$ were calculated using numerical simulation of the geometry (\textit{COMSOL} v5.3 (COMSOL Inc., 2017) and \textit{Sonnet} v16.54 (Sonnet Software Inc., 2017)).
    The contribution of the capacitances $C_{\rm j}$ and $C_{\rm g}$ are expected to be small compared to $C_{\rm s}$.
    The impedances of these (parallel) capacitances are much larger than the typical impedances of the other circuit elements ($L_{\rm j}$ or $R_{\rm sg}$ for example).
    \item Additionally, $L_{\rm g}$ is swept between two extreme values given by our simulations representing a range of possible kinetic inductance values for NbTiN, the superconductor used in our leads.
    This gives the error band shown in Supplementary Figures \ref{fig:figure3_bands} and \ref{fig:figure4_bands}.
\end{itemize}

With this, we are left with three free parameters: $L_{\rm j}$, $R_{\rm sg}$, $C_{\rm s}$.
These are determined from fitting the model to the microwave response of the final device as a function of applied gate voltage $V_{\rm g}$.
In broad terms, $L_{\rm j}$ sets the device resonance frequency, $R_{\rm sg}$ sets the internal quality factor (or loss rate) while $C_{\rm s}$ sets the external quality factor (or coupling).
We note also that points around $V_{\rm g} = V_{\rm CNP}$ fall into a very undercoupled cavity regime, making the resonance peak visibility very low in some cases.
This results in some of our fits not converging to the measured curve and producing absurd results.
Since some of these peaks are not clearly fittable given the measured background, we have opted to reject these few low visibility traces from the final fitted parameter plots.

\section{Feasibility of a Graphene JJ transmon qubit}\label{sec:feasability}
\noindent In this section we provide an additional discussion on the feasibility of a graphene based transmon qubit.

We first consider the device as presented in the main text.
To calculate the anharmonicity of this device we use techniques from the black box quantization method \cite{nigg_blackbox_2012}.
According to this method, the value of the anharmonicity $\alpha$ is then given by
\begin{equation}
\alpha = \frac{2e^2}{L_{\rm j}\omega_0^2(\textrm{Im}(Y'(\omega_0)))},
\end{equation}
where $L_{\rm j}$ is the Josephson inductance of the junction, $\omega_0$ is the resonant frequency of the circuit, $Y$ is the admittance of the circuit seen from the junction terminals (including its own admittance) and $Y'$ its derivative with respect to frequency.
The resonance frequency $\omega_0$ then corresponds to the condition $\textrm{Im}Y(\omega_0) = 0$ and the derivative at this point $Y'(\omega_0)$ can be computed.

As can be seen in Supplementary Figure \ref{fig:anharm1}, the calculated anharmonicity for our main device is always smaller than the measured linewidth.
Therefore it does not qualify as a qubit in its current state.

\section{Design scenario A -- Measured graphene junction in fixed frequency transmon}\label{sec:scenA}

While our device is not immediately a qubit, some improvements are possible.
Most notably, the junction inductance is diluted by the cavity inductance, resulting in a low participation ratio in the total circuit inductance.
We can therefore pose the question of what would the performance of a transmon be that contained only our graphene Josephson junction as its inductive element.
This circuit is shown in the inset in Supplementary Figure \ref{fig:anharm2}a and consists of the junction in parallel with a shunt capacitor $C_{\rm q}$.
The value of this capacitance is set by the requirement that the frequency of the transmon be $\omega_0 = 2\pi\cdot\SI{6}{GHz}$.
Given the measured values of $L_{\rm j}$ as a function of applied gate voltage, we can then obtain the anharmonicity as:
\begin{equation}
\alpha = \frac{e^2}{2C_{\rm q}}.
\end{equation}
The result is shown in Supplementary Figure \ref{fig:anharm2}a along with the projected linewidth of the device $\Gamma = (R_{\rm sg}C_{\rm q})^{-1}$.
Although the situation is improved in this case, the anharmonicity is still substantially lower than the calculated linewidth.
This is due to the fact that we are using a rather wide junction with a somewhat high critical current value and, therefore, a low inductance value.
To keep the frequency at the chosen $\omega_0 = 2\pi\cdot\SI{6}{GHz}$, the necessary capacitance is then too large to make a qubit.
This could be resolved by making our junction narrower, hence increasing its inductance, as we shall see below.

\section{Design Scenario B -- Adjusted width graphene junction in fixed frequency and anharmonicity transmon}\label{sec:scenB}

In this case we consider the same circuit as in the previous case.
Now, however, we fix the capacitance so that the anharmonicity $\alpha = \SI{100}{MHz}$.
This sets the value of our capacitance $C_{\rm q}\simeq\SI{0.2}{pF}$.
Since we also keep the requirement that $\omega_0 = 2\pi\cdot\SI{6}{GHz}$, our junction inductance is fixed to a value of $L_{\rm j} = (\omega_0^2C_{\rm q})^{-1} \simeq \SI{3.5}{nH}$.  
Given these requirements and the measured values of inductance for our device, we can deduce what junction width would be necessary at each gate voltage $V_{\rm g}$ to produce the required inductance.

Here we make the assumption that both $L_{\rm j}$ and $R_{\rm sg}$ scale with the inverse of the junction width, i.e., approximately as $\propto W^{-1}$.
This should be the case for $L_{\rm j}$ since $L_{\rm j} \propto = I_{\rm c}^{-1} \propto R_{\rm n} \propto W^{-1}$ since the $I_{\rm c}R_{\rm n}$ product in a ballistic junction is constant \cite{titov_josephson_2006}.
$R_{\rm sg}$ does not necessarily have to scale as $R_{\rm n}$.
It does, however, depend on the number of conduction channels available and on the graphene proximity gap.
The number of channels should scale linearly with the width of the junction while the proximity gap should increase as high transverse momentum channels are suppressed.
It is therefore reasonable to assume that $R_{\rm sg}$ scales at least as fast as $L_{\rm j}$.

With these assumptions we can then calculate the required width and expected linewidth shown in Supplementary Figure \ref{fig:anharm3}.
In this case there is an ample range of gate voltages that comply with the condition $\Gamma<\alpha$.
The required junction widths are always above \SI{100}{nm}, a limit that is within reach of state of the art fabrication techniques.
It is on this basis that we propose that it is feasible to construct a graphene based transmon qubit.

\section{Hysteresis of the junction switching current}\label{sec:hysteresis}
\noindent The observed hysteresis in the switching current of our devices (see Figure 2a of main text, and Supplementary Figure \ref{fig:repeatdev}a) could have various origins.
A valid estimation of the relevant Stewart-McCumber parameter\cite{tinkham_introduction_1996}, $\beta_{\rm C}=2\pi I_{\rm c}R^2C/\Phi_0$, is not straightforward because there is always the question of how much capacitance of the leads going to the junction should be included.
In principle, for example in DC measurements, even a portion of the wires going up the cryostat could be arguably relevant, up to a point where the inductance of these wires ``chokes'' the capacitance contribution.

We here discuss several estimates of possible relevant capacitances that could enter into $\beta_{\rm C}$, where we assume a typical $R=\SI{50}{\Omega}$ and $I_{\rm s}=\SI{5}{\micro A}$.
First, we note that the ``geometric'' capacitance of a parallel plate capacitor formed between the superconducting leads across the BN/G/BN stack yields a negligible value on the order of a few tens of atto Farads.
More important is the ``local'' stray capacitance of the junction which we have simulated in \textit{COMSOL} v5.3 (COMSOL Inc., 2017) and \textit{Sonnet} v16.54 (Sonnet Software Inc., 2017).
If we include the leads up to a distance of \SI{5}{\mu m} from the junction, the relevant $C=\SI{2}{fF}$ and $\beta_{\rm C}=0.08$.
We also simulated the capacitance of the leads that go from the junction to the surrounding ground plane and to the CPW cavity, giving $C=\SI{6.7}{fF}$ and $\beta_{\rm C}=0.25$.
Of course, there is also likely a relevant capacitance contribution from the center conductor of the CPW to ground.
For this, we can make a rough estimate of the total CPW center conductor capacitance of \SI{909}{fF} and a resulting $\beta_{\rm C}=35$, reaching far into the underdamped regime.
Finally, one could also include the shunt capacitor of \SI{27}{pF}, which would give $\beta_{\rm C}>1000$. 
The last two are likely not completely relevant, since at the Josephson frequency associatated with the finite bias state of the junction ($\omega_{\rm P}=\sqrt{2\pi I_{\rm c}/(\Phi_0 C)}=\SI{24}{GHz}$), the shunt capacitor will not charge through the inductance of the center wire of the cavity. 
More likely, the relevant $\beta_{\rm C}$ includes some reasonable contribution of the CPW capacitance: for example, assuming $C = C_{\rm CPW} / 10 = \SI{90}{fF}$ would give a $\beta_{\rm C} = 3.4$
In addition to these damping effects, self-heating effects inside the SNS junction could further contribute to a hysteretic IVC \cite{courtois_origin_2008,borzenets_phonon_2013}.

\section{Simulation of sub-gap density of states}\label{sec:subgap}
\noindent To gain further insight into the underlying mechanisms of our junction, we model the density of states (DOS) of a gJJ similar to our device with the software package \textit{Kwant} v1.3 \cite{groth_kwant_2014}.
The relevant energies to consider are the bulk superconducting pairing potential $\Delta$ and the effective round-trip time of the Cooper pairs inside the junction, the Thouless energy $E_{\rm th}=\hbar v_{\rm F}/L$.
From the critical temperature of our NbTiN leads (see Supplementary Figure \ref{fig:critical_temp}) we estimate\cite{tinkham_introduction_1996} $\Delta=1.764k_{\rm B} T_{\rm c}\approx\SI{2}{meV}$.
Our device is then placed in the intermediate to long regime, $\Delta/E_{\rm th}\approx1.52>1$.

The modelled system consists of a discretized 2D honeycomb lattice with infinite boundary conditions in y-direction.
The superconducting areas are implemented by setting the pairing potential of these regions to a finite value, effectively making the graphene itself superconducting.
For the simulation shown we assume full SN coupling, corresponding to a contact transparency $Tr=1$.
The simulated system size was $L_{\rm N}=60$ and $L_{\rm SC}=300$ (both in units of the graphene lattice constant $a=\SI{0.214}{nm}$), while we adjusted the pairing potential such that the junction is in the intermediate regime, i.e. $L_{\rm N}/\xi=\Delta/E_{\rm th} = 1.52$.
The dispersion is obtained by solving the eigenvalue problem of the Hamiltonian discretized onto the implemented system and plotting the energy values as a function of transverse momentum $k_\parallel$ (see Supplementary Figure \ref{fig:subgap_dos}).

As expected, there are several Andreev Bound States (ABS) hosted below the bulk gap, significantly reducing $\Delta_{\rm ind}<\Delta_{\rm bulk}$ and opening possible dissipation channels for RF excitations.
As the chemical potential $\mu\gg\Delta$, the subgap states do not change much with doping, in agreement with the relatively flat $R_{\rm sg}$ in Figure 4b of the main text.
In two-dimensional JJs, the aspect ratio can also play a non-negligible role, as there can be a second effective Thouless energy related to the transverse length, or width of the junction, $E_{\rm th}^\parallel=\hbar v_{\rm F}/W_{\rm N}$.
Hence, as the aspect ratio increases, the DOS below the bulk gap can rise significantly.
Alternatively, one can understand this via the subgap dispersion:
ABS with lowest energies are those exhibiting large transverse momentum because their effective path length is longer.
The wider the junction, the longer the maximum direct paths across it become, thus the increase in subgap DOS.
With $W_{\rm N}/L_{\rm N}\approx10$, this is a contributing factor in our device.

Note that this discussion gets more complicated when considering the contact interfaces between the normal and superconducting parts, as for reduced contact transparencies the subgap states are even further pushed towards zero energy.

We confirm the validity of our simulation by calculating the energies of both infinite and finite systems for various scaling factors.
The infinite system is the limit of the finite system with aspect ratio $L_{\rm N} \ll W_{\rm N}$.
For a very narrow gJJ (lateral extension comparable or equal to distance between superconducting contacts), the DOS is much lower below the bulk gap compared to a very wide junction.
The reason for this is the much higher level spacing for a narrow system that pushes additional states above the gap.
Hence, to obtain a SNS system with hard and large induced gap, the normal part should be as narrow and short as possible.

We note that these peaks are not directly visible in our measurements, since instead of measuring the voltage drop across a current-biased JJ they require spectroscopy of the DOS via a tunnel probe, such as in Pillet \emph{et al.} or Bretheau \emph{et al.} \cite{pillet_andreev_2010a,bretheau_tunnelling_2017a}.

\end{document}